\documentclass[a4paper,fleqn,usenatbib]{mnras}

\usepackage{times}

\usepackage[T1]{fontenc}
\usepackage{ae,aecompl}

\usepackage{graphicx}	
\usepackage{amsmath}	
\usepackage{amssymb}	
\usepackage{tabularx}   
\usepackage{pgf}        

\newcommand{\D}[1]{\ensuremath{\text{d}#1}}

\newcommand{\B}[1]{\bmath{#1}}

\newcommand\Eqn[1]     {Eq.\,(\ref{#1})}

\def\mnras{{Mon.~ Not.~ R.~ Astron.~ Soc.~}}

\def\prd{{Phys.~ Rev.~ D.~}}

\def\apj{{Astrophys.~ J.~}}
\def\apjs{{Astrophys.~ J.~ Suppl.~}}
\def\apjl{{Astrophys.~ J.~ Lett.~}}

\def\mnras{{MNRAS}}

\def\prd{{PRD}}

\def\apj{{ApJ}}
\def\apjs{{ApJS}}
\def\apjl{{ApJL}}
\def\aap{{A\&A}}

\def\physrep{{Phys.~ Rep.~}}
\def\jcap{Journal of Cosmology and Astro-Particle Physics}

\newcommand{\be}{\begin{equation}}
\newcommand{\ee}{\end{equation}}
\newcommand{\ba}{\begin{eqnarray}}
\newcommand{\ea}{\end{eqnarray}}

\def\pp1{{\prime}}
\def\pp2{{\prime\prime}}

\def\2D{{\rm 2D}}

\def\bx{\bmath{x}}
\def\br{\bmath{r}}

\def\bk{\bmath{k}}

\def\1Loop{{\rm 1Loop}}

\def\Mpc{\, h^{-1}{\rm Mpc}}

\def\Gpccube{\, h^{-3} \, {\rm Gpc}^3}
\def\kMpc{\, h \, {\rm Mpc}^{-1}}

\def\dk{{\rm d}^3{\bf k}}

\def\nbar{\bar{n}}

\def\la{\mathrel{\mathpalette\fun <}}
\def\ga{\mathrel{\mathpalette\fun >}}
\def\fun#1#2{\lower3.6pt\vbox{\baselineskip0pt\lineskip.9pt
        \ialign{$\mathsurround=0pt#1\hfill##\hfil$\crcr#2\crcr\sim\crcr}}}

\def\Euler{{\rm e}}

\title[Cosmology with Phase Statistics]{Cosmology with Phase Statistics: Parameter Forecasts and
 Detectability of BAO}

\author[A. Eggemeier \& R. E. Smith]{
Alexander Eggemeier,$^{1}$\thanks{E-mail: a.eggemeier@sussex.ac.uk}
and Robert E. Smith$^{1}$
\\
$^{1}$Astronomy Centre, University of Sussex, Falmer, Brighton, BN1 9QH, UK\\
}

\date{Accepted XXX. Received YYY; in original form ZZZ}

\pubyear{2016}

\begin{document}
\label{firstpage}
\pagerange{\pageref{firstpage}--\pageref{lastpage}}
\maketitle

\begin{abstract}
  We consider an alternative to conventional three-point statistics such as the bispectrum, which is purely
  based on the Fourier phases of the density field: the line correlation function. This statistic directly
  probes the non-linear clustering regime and contains information highly complementary to that contained in the
  power spectrum. In this work, we determine, for the first time, its potential to constrain cosmological
  parameters and detect baryon acoustic oscillations (hereafter BAOs). We show how to compute the line
  correlation function for a discrete sampled set of tracers that follow a local Lagrangian biasing scheme and
  demonstrate how it breaks the degeneracy between the amplitude of density fluctuations and the bias parameters
  of the model. We then derive analytic expressions for its covariance and show that it can be written as a sum
  of a Gaussian piece plus non-Gaussian corrections. We compare our predictions with a large ensemble of
  $N$-body simulations and confirm that BAOs do indeed modulate the signal of the line correlation function for
  scales $50$--$100\,\Mpc$, and that the characteristic S-shape feature would be detectable in upcoming
  Stage IV surveys at the level of $\sim4\sigma$. We then focus on the cosmological information content and
  compute Fisher forecasts for an idealized Stage III galaxy redshift survey of volume $V\sim 10\, \Gpccube$ and
  out to $z=1$. We show that, combining the line correlation function with the galaxy power spectrum and a
  Planck-like microwave background survey, yields improvements up to a factor of two for parameters such as
  $\sigma_8$, $b_1$ and $b_2$, compared to using only the two-point information alone.
\end{abstract}






\section{Introduction}
\label{sec:introduction}
The clustering of galaxies in our Universe is commonly analyzed with one of the two simple statistical measures:
the two-point correlation function or its Fourier space analogue, the power spectrum. Measurements of these
statistics in the last decade led to the astonishing detection of baryon acoustic oscillations (hereafter BAO)
in the clustering pattern of galaxies \citep{Eisenstein2005, Tegmark2006} and subsequently to the increasingly
precise constraints on our cosmological models which have confirmed the accelerated expansion rate of the
present day Universe \citep[see e.g.][for a combination of two-point measurements from the final BOSS data
release]{Alam2016}.

Over the next decades, further improvements are expected through developing theoretical advances that will
enable us to extract information from deeper in the non-linear regime and through surveying larger volumes of
space and so beating down the sample variance errors. However, the former is very challenging, especially as the
scales involved approach shell crossing and the realm of uncertain baryonic physics effects takes hold, while
the latter will come to a halt once surveys approach the cosmic variance limit. The path forward thus inevitably
turns the question: what else should we measure to improve our understanding of the Universe? Higher-order
statistics such as the three-point function, or equivalently the bispectrum, are obvious answers. For several
decades the advantages of these measures have been known: the configuration dependence of the bispectrum allows
us to break degeneracies present in the power spectrum, making it efficient in constraining galaxy bias
\citep{Fry1994, Matarrese1997}; it was also shown to place improved constraints on other cosmological parameters
\citep{Sefusatti2006}. Unfortunately, a wider application of these measures has been impeded by the slow
development of improved theoretical modelling of these statistics and the challenge of accurate covariance
matrices, which require very large sets of mock galaxy catalogues \citep[see][for recent measurements of
three-point statistics, though]{Gil-Marin2016, Slepian2016}. For that reason a number of simpler statistics have
been proposed, which only measure a subset of the available three-point information and are designed to
constrain certain parameters. These are for instance the bias estimators defined in \citet{Pollack2014} and
\citet{Schmittfull2015} as well as the position dependent power spectrum \citep{Chiang2014} and the related
skew-spectra \citep{Munshi2010}, which measure squeezed limits of higher-order statistics. Another idea that
aims to compress bispectrum information into only a small number of modes was first explored in
\citet{Regan2012}.

In this paper, we follow yet a different approach. As the power spectrum depends only on the squared amplitude
of the Fourier mode, it is insensitive to the phase of the mode. Consequently, a measure purely based on these
phases will strictly probe information that is not already contained in the power spectrum
\citep{Wattsetal2003}. Provided the density field is Gaussian, the phases remain random and a complete
statistical description can be given in terms of the power spectrum. After initial perturbations start growing
under the influence of gravity, though, correlations among the phases emerge, indicating a flow of information
into higher-order moments of the density field. Phase information is therefore a direct probe of the non-linear
regime and an appropriate measure for non-Gaussian information in the density field.

One phase based statistic that has emerged in recent times is the `line correlation function' (LCF)
\citet{Obreschkow2013}. It is defined as the three-point correlation function of the phases of the density
field, whose three points are equally spaced on a line, each separated by a scale $r$. In \citet{Obreschkow2013}
and \citet{Eggemeier2015} it was shown that the LCF provides a useful measure for quantifying the cosmic web,
being able to differentiate between elongated, filamentary structures and node like structures \citep[see
also][]{Alpaslan2014}. \citet{Obreschkow2013} also showed that the LCF could be used to differentiate between
cold and warm dark matter scenarios. \citet{Wolstenhulme2015} revealed the LCF's relation to conventional
statistical quantities and showed that, at lowest order in standard perturbation theory (hereafter SPT), it can
be expressed as a combination of bispectrum and power spectra, but that it in principle should contain
information from cumulants of even higher order. Several further developments were made by
\citet{Eggemeier2015}, among which was the quantification of the effects of redshift space distortions on the
LCF.

Even though the LCF is closely related to the bispectrum, it should be expected to exhibit different
dependencies on cosmological parameters and also different covariance properties. Furthermore, it offers the
possibility to measure in principal even higher-order information with just a three-point function. All that
makes the LCF an interesting alternative to the standard methods, which is worth investigating more
closely. This paper thus aims to understand the response of the LCF to variations in the underlying cosmological
models and to quantify the signal-to-noise of the estimates. Coupling together these quantities we aim to assess
the cosmological parameter sensitivity of the LCF for upcoming galaxy redshift surveys that can be broadly
classed as Stage III and Stage IV, in the language of the Dark Energy Task Force \citep{DETF2006}.

The paper is broke down as follows: in \S\ref{sec:predictions} we provide a brief review of the LCF, we show how
to compute the galaxy LCF and develop an analytic derivation of its covariance matrix. These theoretical
predictions are then confronted with measurements from a large ensemble of N-body simulations in
Sec.~\ref{sec:measurements}, before we move on to discuss the parameter sensitivity and error forecasts in
Sec.~\ref{sec:cosmoinfo}, supported by a number of N-body simulations with varying
cosmologies. Sec.~\ref{sec:discussion} finally gives our conclusions.


\section{Predictions from Theory}
\label{sec:predictions}


\subsection{The matter line correlation function}
\label{sec:definitions}


Consider a large volume of space within which is the realisation of a statistical homogeneous and isotropic
random field. For a given scale $r$, the LCF of matter fluctuations can be defined as \citep[for details
see][]{Obreschkow2013,Wolstenhulme2015}:
\be \ell_{\rm m}(r) \equiv
V^3\,\left(\frac{r^3}{V}\right)^{3/2}
\left<\frac{}{}\epsilon_{r}(\bx)\epsilon_{r}(\bx+\br)\epsilon_r(\bx-\br)\right> \label{eq:l1}
\ee
where the first factor is a volume regularisation term, with $V$ being the volume of the survey and
$\epsilon_r(\bx)$ is the real space phase field smoothed on scale $r$. The Fourier transform of the smoothed
phase field can be written:
\be \epsilon_r(\bx) = \int \frac{\dk}{(2\pi)^3} \epsilon_{\bk} 
\Euler^{i\bk\cdot\bx} W(k|r) \, \label{eq:epsi}\ee
where the window function is a spherical top-hat in $k$-space: $W(k|r)=\Theta(1-kr/2\pi)$ and with $\Theta(x)$
being the Heaviside function. The phase field can now be readily defined as $\epsilon(\bk)\equiv
\delta(\bk)/|\delta(\bk)|$, with $\delta(\bk)$ being the Fourier transform of the overdensity field. Note that
the angle brackets in \Eqn{eq:l1} indicate an averaging over an ensemble of random fields. Under the assumption
of Ergodicity of the fields this becomes an average over volume and orientation of the direction vector of the
line $\hat{\br}$ at each point in space.

On substitution of \Eqn{eq:epsi} into \Eqn{eq:l1} we find that the line correlation can also be written:
\begin{align}
  \label{eq:predictions.ldef}
  \ell_{\rm m}(r) = &\,\frac{V^3}{(2\pi)^{9}}\,\left(\frac{r^3}{V}\right)^{3/2} \hspace{-1.3em}
  \underset{{|\B{k}_1|,|\B{k}_2|,|\B{k}_3| \leq 2\pi/r}}{\iiint}
\hspace{-0.5em}\D{^3\B{k}_1}\,\D{^3\B{k}_2}\,\D{^3\B{k}_3}\,\nonumber \\[0.2em]
&\times\,\left<\epsilon_{\B{k}_1}\,\epsilon_{\B{k}_2}\,\epsilon_{\B{k}_3}\right>\,
\int \frac{\D{\hat{\br}}}{4\pi}  \text{e}^{i[\B{k}_1 \cdot \B{x}+\B{k}_2 \cdot (\B{x}+\B{r})+\B{k}_3 \cdot (\B{x}-\B{r})]}\,, 
\end{align}
with the solid angle element $\D{\hat{\B{r}}} \equiv \sin{\vartheta}\,\D{\vartheta}\,\D{\varphi}$. In order to
proceed further one has to compute the ensemble average of the three phase factors,
\begin{align}
  \left<\epsilon_{\B{k}_1}\epsilon_{\B{k}_2}\epsilon_{\B{k}_3}\right>
=\left<\exp{\left[i\left(\theta_{\B{k}_1}+\theta_{\B{k}_2}+\theta_{\B{k}_3}\right)\right]}\right> \ .
\end{align}
This expression can be evaluated by means of the joint probability density function (PDF) of Fourier phases,
${\cal P}(\{\theta_{\B{k}}\})$, which was derived in \citet{Matsubara2003, Matsubara2006}. As is detailed in
App.~\ref{sec:PDF}, for weakly non-Gaussian fields the PDF can be expanded in an Edgeworth series of higher
order correlators (in Fourier space poly-spectra) and using this result, one finds that to lowest order
\citep{Wolstenhulme2015}:
\begin{align}
  \label{eq:predictions.phase_average}
  \left<\epsilon_{\B{k}_1}\,\epsilon_{\B{k}_2}\,\epsilon_{\B{k}_3}\right>
  \approx &\,\frac{(2\pi)^3}{V} \left(\frac{\sqrt{\pi}}{2}\right)^3\,
  \frac{B(\B{k}_1,\B{k}_2,\B{k}_3)}{\sqrt{V\,P(\B{k}_1)\,P(\B{k}_2)\,P(\B{k}_3)}}\,
  \nonumber\\ &\times\,\delta_D(\B{k}_1+\B{k}_2+\B{k}_3)\,.
\end{align}
where the Dirac delta function $\delta_D$ appears in the above expression as a consequence of statistical
homogeneity of the phase field, and where $P$ and $B$ are the power spectrum and bispectrum of the matter field,
respectively. These are defined:
\begin{align}
  \label{eq:predictions.Pdef}
  \left<\delta_{\B{k}}\,\delta_{\B{k}'}\right> &\equiv (2\pi)^3\delta_D(\B{k}+\B{k}')\,P(\B{k})\,\, ;\\
  \label{eq:predctions.Bdef}
  \left<\delta_{\B{k}_1}\,\delta_{\B{k}_2}\,\delta_{\B{k}_3}\right> &\equiv (2\pi)^3\delta_D(\B{k}_1+\B{k}_2+\B{k}_3)
\,B(\B{k}_1,\B{k}_2,\B{k}_3)\,
  \,.
\end{align}
The explicit dependence of Eq.~(\ref{eq:predictions.phase_average}) on the volume and hence the suppression of
phase correlations in bigger surveys might appear surprising. However, with increasing volume the density field
will contain a greater number of halos and thus peaks in density. As has been shown in \citet{Hikage2004}, the
more peaks of comparable heights are enclosed in the sampling volume, the more the distribution of the phase sum
will approach a uniform value. If this distribution becomes uniform, phase correlations are consequently
diluted.

Finally on inserting \Eqn{eq:predictions.phase_average} into \Eqn{eq:predictions.ldef} and integrating over
$\bk_1$ we find that at leading order, the line correlation can be written as an integral of the form:
\begin{align}
  \ell_{ m}(r) \approx &\,r^{9/2} \left(\frac{\sqrt{\pi}}{2}\right)^3
  \underset{{|\B{k}_1|,|\B{k}_2|,|\B{k}_1+\B{k}_2| \leq 2\pi/r}}{\iint}
  \frac{\D{^3\B{k}_1}}{(2\pi)^3}\,\frac{\D{^3\B{k}_2}}{(2\pi)^3}\,\nonumber \\[0.2em] & \times\,
  \frac{B(\B{k}_1,\B{k}_2,\B{k}_1+\B{k}_2)}{\sqrt{\,P(\B{k}_1)\,P(\B{k}_2)\,P(\B{k}_1+\B{k}_2)}} \int
  \frac{d\hat{\br}}{4\pi} \Euler^{i(\B{k}_1-\bk_2)\cdot\br} \ ,
\end{align}
where after computing the integral we made the following relabelings $\bk_2\rightarrow\bk_1$ and
$\bk_3\rightarrow\bk_2$. Finally, on computing the average over all orientations of $\br$ we arrive at the
result:
\begin{align}
\ell_{\rm m}(r) \simeq &\,
\bigg(\frac{r}{4\pi}\bigg)^{9/2}
\hspace*{-0.5em}\underset{{|\B{k}_1|,|\B{k}_2|,|\B{k}_1+\B{k}_2|
\leq 2\pi/r}}{\iint}\hspace{-0.5em} \D{^3\B{k}_1}\,\D{^3\B{k}_2} \nonumber \\[0.5em] &
\times\,\frac{B(\B{k}_1,\B{k}_2,\B{k}_1+\B{k}_2)}{\sqrt{P(\B{k}_1)\,P(\B{k}_2)\,P(\B{k}_1+\B{k}_2)}}\,
j_0\left(\left|\B{k}_1-\B{k}_2\right| r\right)\,, \label{eq:ilcf-perturbative}
\end{align}
where $j_0(x)=\sin x/x$ is the zeroth order spherical Bessel function.

\subsection{LCF from standard perturbation theory}

On applying nonlinear SPT to the fluid equations for primordial Gaussian matter fluctuations in an expanding
universe, one finds that at lowest order (tree-level) the bispectrum can be expressed in terms of the linear
power spectrum $P$ and a mode coupling kernel $F_2$ as \citep{Fry1984,Bernardeauetal2002}:
\begin{align}
  \label{eq:predictions.BPT}
  B(\B{k}_1,\B{k}_2,\B{k}_3) =
  2\,F_2(\B{k}_1,\B{k}_2)\,P_1\,P_2 + \text{cyc.}\,,
\end{align}
where in the above and in what follows we will use the short-hand notation \mbox{$P_i\equiv P(k_i)$} and
\mbox{$P_{ij}\equiv P(|\bk_i-\bk_j|)$} and the power spectra are understood to be those obtained from linear
theory. The mode coupling kernel $F_2$ reads:
\begin{align}
  \label{eq:predictions.F2}
  F_2(\B{k}_1,\B{k}_2) = \frac{5}{7}+\frac{\hat{\B{k}}_1 \cdot \hat{\B{k}}_2}{2}\left(\frac{k_1}{k_2} +
    \frac{k_2}{k_1}\right)+\frac{2}{7}\left(\hat{\B{k}}_1 \cdot \hat{\B{k}}_2\right)^2\,,
\end{align}
dropping its very weak dependence on the cosmological parameters \citep{Bernardeauetal2002}. On using again
\Eqn{eq:predictions.phase_average} and inserting Eq.~(\ref{eq:predictions.BPT}) in this expression one can now
choose to perform the integrations over the Dirac delta function so that the matter LCF at tree level can be
expressed as \citep{Wolstenhulme2015}:
\begin{align}
  \ell_{\rm m}(r) = & \, 2\left(\frac{r}{4\pi}\right)^{9/2} \hspace{-0.5em}
  \underset{{|\B{k}_1|,|\B{k}_2|,|\B{k}_1+\B{k}_2| \leq 2\pi/r}}{\iint}
  \hspace{-0.5em}\D{^3\B{k}_1}\,\D{^3\B{k}_2}\,F_2(\B{k}_1,\B{k}_2)\nonumber
  \\[0.2em]  &\hspace{-0.2cm}\times\,\,\sqrt{\frac{P_1\,P_2}{P_{12}}}\,
  \Big[j_0{\big(\left|\B{k}_2-\B{k}_1\right|r\big)}+
  2j_0{\big(\left|\B{k}_1+2\B{k}_2\right|r\big)}\Big]\,. \label{eq:predictions.lPT}
\end{align}
While the above expression appears to require a 6-D integration, in fact it only requires a 3-D
integration. This owes to the fact that all terms can be expressed in terms of the magnitudes of $\bk_1$ and
$\bk_2$ and the angle between these vectors. Note that in deriving the above result we have assumed that for the
denominator of Eq.~(\ref{eq:predictions.phase_average}) one-loop corrections to the power spectra are of
negligible importance. As we will show in Sec.~\ref{sec:measurements}, for the range of scales that we will
consider, this assumption appears to be reasonably accurate.

Finally, regarding the cosmological sensitivity of the matter LCF, we see from \Eqn{eq:predictions.lPT} that at
tree-level the LCF is built from integrals over products of the linear matter power spectrum on different
scales. Thus the cosmological dependence derives entirely from specifying the dependence of the linear power on
cosmology. This can best be achieved with Boltzmann codes like \texttt{CAMB} \citep{Lewis2000}.


\subsection{The galaxy line correlation function}
\label{sec:galaxyLCF}

We next turn to the issue of predicting the galaxy LCF (hereafter GLCF). There are two sources of
non-Guassianity that can contribute here, one stems from the relation between the galaxy over density field
$\delta_g$ and that of the matter -- otherwise known as galaxy bias. The second from the fact that galaxies are
a point sampled process and with the usual assumption that they share the same limiting properties for rare
occupancy of micro-cells as a Poisson sampling process.

Firstly, for the galaxy bias, if we assume that the density field in real space obeys the cosmological
principle, in that it is statistically homogeneous and isotropic on sufficiently large scales, one can argue
that in this regime the relation between $\delta_g$ and $\delta$ should become linear
\citep{Fry1993,Smithetal2007}. For this case the bias simply drops out of the equations for the GLCF,
i.e. $\epsilon_g(\B{x}) = \epsilon(\B{x})$ and the galaxy and dark matter LCFs fully coincide.

On smaller scales non-linear terms enter the bias relation \citep{Fry1993,Smithetal2007}. Furthermore, if the
relation does not depend on the present day local density but on the density of the initial patches, then
non-local bias terms can contribute \citep{Catelanetal2000,McDonald2009,Chan2012,Baldauf2012}. For that reason
we employ the Eulerian non-linear, non-local bias model that was proposed by \citet{McDonald2009}. In their
model non-locality is introduced via a term which is quadratic in the tidal tensor $s_{ij}(\B{x}) \equiv
\left[\partial_i\partial_j\nabla^{-2}-\frac{1}{3}\delta^K_{ij}\right] \delta(\B{x})$, such that the galaxy
overdensity may be written as
\begin{align}
  \label{eq:predictions.bias}
  \delta_g(\B{x}) =\,&b_1\,\delta(\B{x}) + \frac{1}{2}b_2\Big[\delta^2(\B{x})-\left<\delta^2(\B{x})\right>\Big]
  \nonumber \\ &+ \frac{1}{2}b_{s^2}\Big[s^2(\B{x})-\left<s^2(\B{x})\right>\Big] +\,\ldots\,,
\end{align}
where $s^2 = s^{ij} s_{ji}$ and the dots indicate terms of even higher orders. The constants $b_1$, $b_2$,
$b_{s^2}$ denote the linear, non-linear and non-local bias terms, respectively, and the terms
$\left<\delta^2\right>$ as well as $\left<s^2\right>$ ensure that $\left<\delta_g\right> = 0$.

%
\begin{figure}
  \centering
  \includegraphics{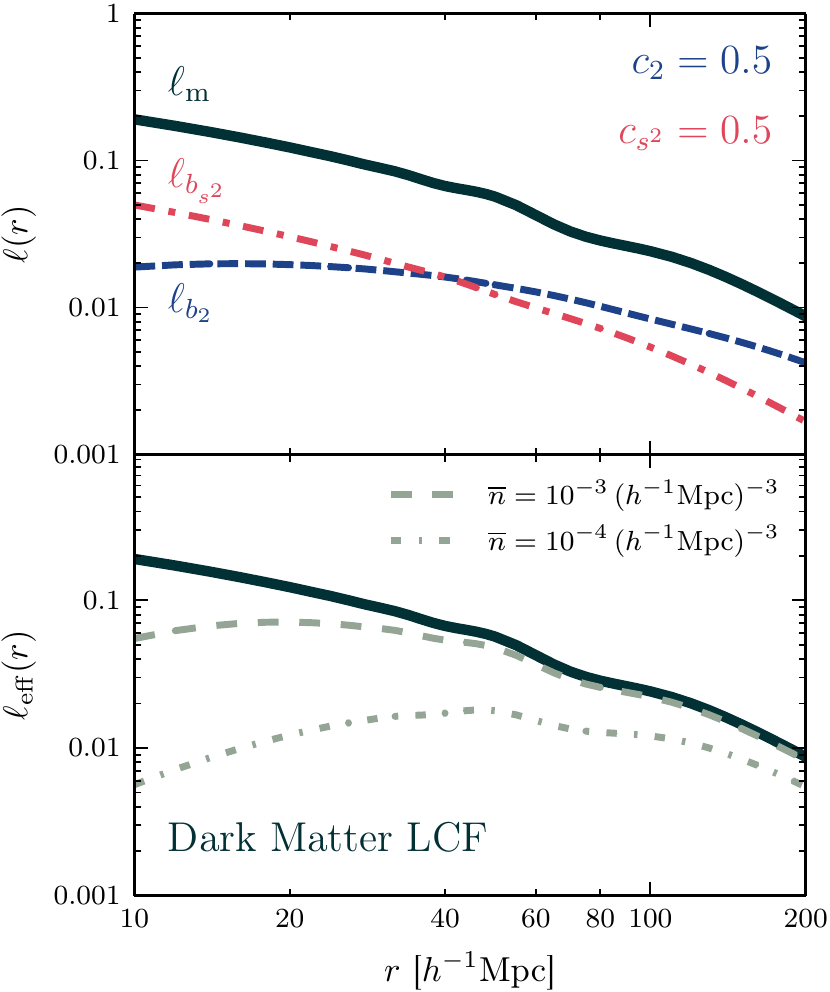}
  \caption{\textit{Upper panel:} Comparison of various contributions to the galaxy LCF with $\ell_{b_2}(r)$ in
    blue and $\ell_{b_{s^2}}(r)$ in red. The coefficients $c_2$ and $c_{s^2}$ are both set to
    $0.5$. \textit{Lower panel:} Suppression effect from shot noise on the dark matter LCF, different line
    styles correspond to different number densities. The thick, black line in both panels is the LCF of the
    underlying matter field, i.e. for negligible shot noise.}
  \label{fig:lgalaxy-shot}
\end{figure}
%

From the Fourier transform of Eq.~(\ref{eq:predictions.bias}) and by using PT we can readily determine the
galaxy bispectrum. As before it can be expressed in terms of the linear dark matter power spectrum:
\begin{align}
    \label{eq:predictions.BgPT}
  B_{\rm g,123}  =  b_1^3\,P_1\,P_2\,\Big[2 F_2(\B{k}_1,\B{k}_2)
    + c_2 + c_{s^2} S_2(\B{k}_1,\B{k}_2)\Big]+ \text{cyc.}\,,
\end{align}
where we used the short-hand notation $B_{\rm g,123}\equiv B_{\rm g}(\bk_1,\bk_2,\bk_3)$ and defined $c_2 \equiv
b_2/b_1$ and $c_{s^2} \equiv b_{s^2}/b_1$, and where we have neglected all higher order terms. We note that the
non-local bias term has introduced an extra configuration dependence, which is encoded in the new kernel
function \citep{McDonald2009}:
\begin{align}
  \label{eq:predictions.S2}
  S_2(\B{k}_1,\B{k}_2) = \left(\hat{\B{k}}_1 \cdot \hat{\B{k}}_2\right)^2 - \frac{1}{3}\,.
\end{align}
Instead of relating the galaxy phase field to the matter one via Eq.~(\ref{eq:predictions.bias}), we observe
that the galaxy phase PDF can be obtained from the original PDF by replacing all spectra in its Edgeworth
expansion with their biased equivalents. Hence, the three-point correlation of galaxy phase factors is the same
as Eq.~(\ref{eq:predictions.phase_average}) with the dark matter quantities replaced by the galaxy bispectrum
and power spectrum. As before we do not consider higher-order corrections to the denominator of
Eq.~(\ref{eq:predictions.phase_average}), which are now also coming from non-linear and non-local bias, and
simply substitute the linear galaxy power spectrum $P_{L,g} = b_1^2\,P_L$. We thus find that the GLCF can be
written to lowest order as the sum of three terms:
\begin{align}
  \label{eq:predictions.lgPTterms}
  \ell_{\rm g}(r) = \ell_{\rm m}(r) + c_2\ell_{b_2}(r) + c_{s^2}\ell_{b_{s^2}}(r)\,,
\end{align}
where each term can be written:
\begin{align}
  \ell_{\alpha}(r) = &\, \left(\frac{r}{4\pi}\right)^{9/2} \hspace{-0.5em}
  \underset{{|\B{k}_1|,|\B{k}_2|,|\B{k}_1+\B{k}_2| \leq 2\pi/r}}{\iint}
  \hspace{-0.5em}\D{^3\B{k}_1}\,\D{^3\B{k}_2}\sqrt{\frac{P_1\,P_2}{P_{12}}}
  \nonumber \\[0.2em]
  &\hspace{-0cm}\times\,\Big[j_0{\big(\left|\B{k}_2-\B{k}_1\right|r\big)}+
  2j_0{\big(\left|\B{k}_1+2\B{k}_2\right|r\big)}\Big]
  \Gamma_{\alpha}(\bk_1,\bk_2)\,, \label{eq:predictions.lgPT}
\end{align}
where the above equation holds for \mbox{$\alpha \in \{{\rm m},b_2,b_s^2\}$} and where
\mbox{$\Gamma_{\alpha}(\bk_1,\bk_2)=\{2F(\bk_1,\bk_2),1,S_2(\bk_1,\bk_2)\}$}.
At this point we draw attention to one of the major advantages of the GLCF over more conventional galaxy
clustering statistics, which is that in the limit of linear bias, i.e. where $b_2 = b_{s^2}= 0 $, we see that
$\ell_{\rm g}(r) = \ell_{\rm m}(r)$. Hence it is a direct probe of the matter distribution, independent of linar
bias. We also notice that in the presence of non-trivial biasing the GLCF is sensitive only to the relative bias
parameters, $c_2 = b_2/b_1$ and $c_{s^2} = b_{s^2}/b_1$.

Figure~\ref{fig:lgalaxy-shot} (upper panel) shows all three terms. The dark matter LCF is indicated by the
(black) thick line, and the dashed and dot-dashed lines represent the contribution of the local and nonlocal
bias terms, respectively, where we set $c_2 = c_{s^2} = 0.5$. To evaluate the various expressions in
Eq.~(\ref{eq:predictions.lgPTterms}) we adopted the same $\Lambda$CDM cosmology as described in
Sec.~\ref{sec:measurements} and generated the corresponding linear power spectrum using \texttt{CAMB}. We notice
the different configuration dependence of all three functions, meaning that in principle it is possible to
determine both $c_2$ and $c_{s^2}$, as well as the amplitude of fluctuations $\sigma_8$ from the LCF
alone. However, the sole discernible difference between $\ell_{\rm m}(r)$ and $\ell_{c_{s^2}}(r)$ are the damped
BAO wiggles in the latter, so we cannot expect the LCF to yield strong constraints on $c_{s^2}$. We stress that
for the bispectrum there will be a remaining degeneracy between $\sigma_8$ and $b_1$ as can be seen from
Eq.~(\ref{eq:predictions.BgPT}), which must either be broken via a joint analysis with the power spectrum or
under the inclusion of a cosmic microwave background (hereafter CMB) measurement.

For the remainder of this work we will additionally assume that bias is local in Lagrangian space, in which case
it can be shown that the non-local bias term is related to the linear one at first order
\citep{Baldauf2012, Chan2012},
\begin{align}
  b_{s^2} = -\frac{4}{7}\left(b_1-1\right)\,.
\end{align}
In this case galaxy biasing is only a function of two free parameters, $b_1$ and $b_2$, and as we will show
these can be efficiently constrained through combination of the LCF with the power spectrum.


\subsection{The effect of shot noise on the GLCF}
\label{sec:shotnoise}

Another additional source of non-Gaussianity that modulates the GLCF is sampling noise. Any real measurement
from a galaxy survey will be compromised by shot noise -- the fact that the matter field has to be reconstructed
from a discrete and finite set of tracers that have been sampled from some underlying field which may be
Gaussian. This effect increases as the density of objects in a given volume decreases and leads to an artificial
enhancement in the clustering strength of galaxies in the power spectrum. In the absence of a selection function
and finite survey geometry, a constant number density of tracers $\overline{n}$, would modulate the true power
spectrum and bispectrum as follows:
\citep{Peebles1980, Matarrese1997},
\begin{align}
  \label{eq:predictions.Pdiscrete}
  P^{\rm d}(k) & = P(k) + \frac{1}{\overline{n}}\,, \\
  \label{eq:predictions.Bdiscrete}
  B^{\rm d}(k_1,k_2,k_3) & = B_{123} + \frac{1}{\overline{n}}\left[P_1+P_2+P_3\right] + \frac{1}{\overline{n}^2}\,,
\end{align}
where the superscript ${\rm d}$ stands for the discrete case. All shot noise terms involving factors of
$1/\overline{n}$ can be subtracted to obtain an unbiased estimate of the true power spectrum or bispectrum,
although they will still contribute to the errors.

To derive the effect of shot noise on the GLCF, we employ the same trick as employed in the last section. We
assume the Fourier modes of the reconstructed matter field follow the same PDF as that of the true matter field,
but with all spectra in the Edgeworth expansion replaced by the corresponding discrete quantities. Accordingly,
using Eqs.~(\ref{eq:predictions.Pdiscrete}) and (\ref{eq:predictions.Bdiscrete}) the three-point phase
correlator estimated from a set of discrete tracers is given by
\begin{align}
  \label{eq:predictions.phasediscrete}
  \left<\epsilon_{\B{k}_1}\epsilon_{\B{k}_2}\epsilon_{\B{k}_3}\right>^{\rm d}
  = &\,\frac{(2\pi)^3}{V}\left(\frac{\sqrt{\pi}}{2}\right)^3
  \sqrt{\frac{\nu_{\text{eff}}(k_1)\,\nu_{\text{eff}}(k_2)\,\nu_{\text{eff}}(k_3)}{V\,P_1\,P_2\,P_3}}
  \nonumber \\ &\times \Bigg[B_{123} +
    \frac{1}{\overline{n}}\Big(P_1+P_2+P_3\Big) +
    \frac{1}{\overline{n}^2}\Bigg] \nonumber \\ &\times
  \delta_D(\B{k}_1+\B{k}_2+\B{k}_3)\,,
\end{align}
where we have defined
\begin{align}
  \nu_{\text{eff}}(k) \equiv
  \frac{\overline{n}\,P(k)}{1+\overline{n}\,P(k)}\,.
\end{align}
This factor encodes the shot noise contamination of each mode and is related to the effective volume of the
survey \citep{Tegmark1997b},
\begin{align}
  \label{eq:predictions.Veff}
  V_{\text{eff}}(k) = \int \D{^3x}
  \left(\frac{\overline{n}(\B{x})\,P(k)}{1+\overline{n}(\B{x})\,P(k)}\right)^2\,,
\end{align}
such that for a constant number density, $\nu_{\text{eff}}(k) =\sqrt{V_{\text{eff}}(k)/V}$. 

The discrete form of the LCF can now be computed by substituting \Eqn{eq:predictions.phasediscrete} into
\Eqn{eq:predictions.ldef} and proceeding as before, whereupon we see that we may write the {\em effective} LCF
as:
\begin{align}
  \label{eq:measurements.leff} 
  \hat{\ell}_{\rm eff}(r) = \hat{\ell}^{\rm d}(r) - \hat{\ell}_{\rm shot}(r)\,,
\end{align}
where the second term on the right-hand-side is a pure shot-noise term which has the form:
\begin{align}
  \label{eq:measurements.lshot}
  \hat{\ell}_{\text{shot}}(r) = &\,8\pi^2\left(\frac{r}{4\pi}\right)^{9/2}\int_{0}^{2\pi/r} \D{k_1}\,k_1^2
  \int_0^{2\pi/r} \D{k_2}\,k_2^2
  \int_{-1}^{\mu_{\text{cut}}} \D{\mu} \nonumber \\
  &\times\,\sqrt{\frac{\nu_{\text{eff}}(k_1)\,\nu_{\text{eff}}(k_2)\,\nu_{\text{eff}}(|\B{k}_1+\B{k}_2|)}
    {\hat{P}_1\,\hat{P}_2\,\hat{P}_{12}}} \nonumber \\
  &\times\,\left[\frac{1}{\overline{n}}\left(\hat{P}_1+\hat{P}_2+\hat{P}_{12}\right)+
    \frac{1}{\overline{n}^2}\right]\,j_0(|\B{k}_2-\B{k}_1|r)\,,
\end{align}
where
\be \mu_{\text{cut}} = \text{min}\left\{1,\,\text{max}\left\{-1,
\left[(2\pi/r)^2-k_1^2-k_2^2\right]/2k_1k_2\right\}\right\}\ \ee
guarantees that $|\B{k}_1+\B{k}_2| \leq 2\pi/r$ and $\hat{P}_{12} \equiv \hat{P}(|\B{k}_1+\B{k}_2|)$. Note that
unlike $P^{\rm d}$ and $B^{\rm d}$ which may be fully corrected for `Poisson-like' shot noise, the LCF can not
(hence the name effective) since the shot noise contribution that enters via the factor $\sqrt{\nu_{\rm
    eff}(k_1)\dots/P^{\rm d}_1\dots}$ in \Eqn{eq:predictions.phasediscrete} can not be fully separated. Hence,
it remains part of the estimator, but note that in the limit of $\nbar P\gg 1$ the effect of shot-noise is
negligible and we fully recover $\ell_{\rm m}$. On the other hand, in the limit of $\nbar P\ll1$, the estimate
is shot noise dominated and the LCF scales as $\propto 1/\sqrt{\nbar}$ multiplied by the three-point
self-correlation function for a k-space top-hat filter function evaluated for a line configuration with the
regularisation factor.

The subtraction of the shot-noise terms as described above leads to a suppression of the GLCF that has to be
taken into account when comparing measurements to model predictions.  Figure~\ref{fig:lgalaxy-shot}, bottom
panel, shows the impact of this suppression effect on the true dark matter LCF (thick, solid line) for two
different number densities. The smallest scales are most heavily affected, which is reasonable as most of the
high $k$-modes are efficiently damped away by $\nu_{\text{eff}}$. Even for a number density of $\overline{n} =
10^{-3}\,(h^{-1}\text{Mpc})^{-3}$ the suppression is significant, ranging from $\sim 70\,\%$ at $10\,h^{-1}$Mpc
to still $\sim 15\,\%$ at the scale of the first BAO bump ($50\,h^{-1}$Mpc). In Sec.~\ref{sec:measurements} we
will confront Eq.~(\ref{eq:predictions.phasediscrete}) with measurements from N-body simulations.

Before moving on, we note that the suppression of the GLCF due to point sampling can be understood in a rather
intuitive way: when reconstructing the matter field, each tracer contributes with a single peak convolved with
some narrow window function. If the density of tracers is decreased further and further, this field will tend to
look like a collection of many separate peaks of nearly equal heights instead of reflecting the true underlying
matter field with its density peaks of various heights and sizes. This leads to a suppression of phase
correlations because the presence of many peaks with comparable heights renders the phase distribution nearly
uniform, as has already been noted above \citep{Hikage2004}.


\section{The covariance of the line correlation}
\label{sec:LCFcovariance}

In order to study how much cosmological information a combined measurement of the LCF and power spectrum
provides, we will need to compute the auto-covariance properties of the LCF and its cross-covariance with the
power spectrum. The aim of this section is to provide analytic expressions for these quantities. Since this is a
rather technical section, we suggest that for those not wishing to plough through the calculations at this stage
they skip \S\ref{sec:autocov} and \ref{sec:crosscov}.


\subsection{Estimators}\label{sec:estimators}

In general we write the full joint covariance matrix of the LCF and power spectrum as:
\begin{align}
  \label{eq:predictions.covdef}
  \mathsf{C}_{ij} \equiv \left<\delta X_i\,\delta X_j\right>\,,
\end{align}
where $\delta X_i = X_i-\left<X_i\right>$ and $X_i$ can either stand for the LCF estimator $\hat{\ell}_i$ at
some radial bin $r_i$, or the power spectrum estimator $\hat{P}_i$ with bin $k_i$. In order to obtain
expressions for the estimators of the theoretical definitions in Eqs.~(\ref{eq:predictions.ldef}) and
(\ref{eq:predictions.Pdef}) we apply the following prescription: we assume that the survey volume is large
enough to encompass many independent patches of the universe and hence replace the ensemble average with an
average over volume. Performing the corresponding integrations, we are able to write the estimator for the LCF
as:
\begin{align}
  \label{eq:predictions.lestimator}
  \hat{\ell}(r) = &\,\frac{V^2}{(2\pi)^{6}}\,\left(\frac{r^3}{V}\right)^{3/2} \hspace{-1.3em}
  \underset{{|\B{k}_1|,|\B{k}_2|,|\B{k}_1+\B{k}_2| \leq 2\pi/r}}{\iint}
  \hspace{-0.5em}\D{^3\B{k}_1}\,\D{^3\B{k}_2}\,\nonumber \\[0.2em]
  &\times\,j_0(|\B{k}_1-\B{k}_2|\,r)\,\epsilon_{\B{k}_1}\,\epsilon_{\B{k}_2}\,\epsilon_{-\B{k}_1-\B{k}_2}\,.
\end{align}
Similarly, for the power spectrum averaged over a bin of with $\Delta k$,
\begin{align}
  \label{eq:predictions.Pestimator}
  \hat{P}(k) = \frac{1}{V}\,\int_k \D{^3q_1} \int_k \D{^3q_2}
  \frac{\delta_D(\B{q}_1+\B{q}_2)}{V_P(k)}\,\delta_{\B{q}_1}\,\delta_{\B{q}_2}\,,
\end{align}
where the integrals run over $|\B{q}| \in \left[k-\Delta k/2, k+\Delta k/2\right]$ and \mbox{$V_P \equiv 4 \pi
  k^2 \Delta k$} is the volume of a spherical shell in Fourier space.

Our main task is then to evaluate ensemble averages of phase factors $\exp{(i\theta_{\B{k}})}$ as well as
combinations of phase factors with amplitudes $A_{\B{k}}$, which are given as integrals over the joint PDF of
Fourier modes ${\cal P}(\{A_{\B{k}},\theta_{\B{k}}\})$. App.~\ref{sec:PDF} demonstrates how this PDF can be
expanded perturbatively in a series containing all higher order spectra, where the order of the contributing
terms can be conveniently labeled by powers of $1/\sqrt{V}$. This expansion is then used to derive all ensemble
averages to lowest order needed for the subsequent computations.


\subsection{Auto-covariance matrix of $\hat{\ell}$}
\label{sec:autocov}

According to Eq.~(\ref{eq:predictions.covdef}), the central quantity for the LCF covariance is the six-point
phase correlator subtracted by the mean,
\begin{align}
 {\cal E}_{\ell\ell} \equiv \left<\epsilon_{\B{k}_1}\,\epsilon_{\B{k}_2}\,\epsilon_{\B{k}_3}
  \,\epsilon_{\B{q}_1}\,\epsilon_{\B{q}_2}\,\epsilon_{\B{q}_3}\right> -
\left<\epsilon_{\B{k}_1}\,\epsilon_{\B{k}_2}\,\epsilon_{\B{k}_3}\right>\, 
  \left<\epsilon_{\B{q}_1}\,\epsilon_{\B{q}_2}\,\epsilon_{\B{q}_3}\right>\,,
\end{align}
where it is implied that $\B{k}_3 = -\B{k}_1-\B{k}_2$ and $\B{q}_3 = -\B{q}_1-\B{q}_2$. Using the cumulant
expansion theorem, we can split this correlator into its various connected pieces as follows,
\begin{align}
  \label{eq:predictions.sixphase_expansion}
  {\cal E}_{\ell\ell} =
  \,&\left<\epsilon_{\B{k}_1}\,\epsilon_{\B{q}_1}\right>\,\left<\epsilon_{\B{k}_2}\,\epsilon_{\B{q}_2}\right>\,
  \left<\epsilon_{\B{k}_3}\,\epsilon_{\B{q}_3}\right> + \text{sym.}(6) \nonumber \\
  &+ \left<\epsilon_{\B{q}_1}\,\epsilon_{\B{k}_2}\,\epsilon_{\B{k}_3}\right>_c\, 
  \left<\epsilon_{\B{k}_1}\,\epsilon_{\B{q}_2}\,\epsilon_{\B{q}_3}\right>_c + \text{sym.}(9) \nonumber \\
  &+ \left<\epsilon_{\B{k}_1}\,\epsilon_{\B{k}_2}\,\epsilon_{\B{q}_1}\,\epsilon_{\B{q}_2}\right>_c\,
  \left<\epsilon_{\B{k}_3}\,\epsilon_{\B{q}_3}\right> + \text{sym.}(9) \nonumber \\
  &+ \left<\epsilon_{\B{k}_1}\,\epsilon_{\B{k}_2}\,\epsilon_{\B{k}_3}
  \,\epsilon_{\B{q}_1}\,\epsilon_{\B{q}_2}\,\epsilon_{\B{q}_3}\right>_c\,,
\end{align}
with $\text{sym.}(n)$ indicating that $n-1$ terms have to be added to symmetrize the corresponding expressions
with respect to the $\B{k}$'s and $\B{q}$'s. Connected correlators consist of all those terms that cannot be
written as a product of two or more connected pieces and for the phase fields they result from the PDF expansion
coefficients where $m=1$ in Eq.~(\ref{eq:PDF.Qn}). As such, each connected phase correlator is not limited to a
single spectrum but a series of all even or odd spectra, which become increasingly suppressed by factors of
volume. The exception is the two-point function of phase factors, where statistical homogeneity dictates that
$\B{k}_1=-\B{k}_2$, in which case $\epsilon_{\B{k}_1}\epsilon_{\B{k}_2} = |\epsilon_{\B{k}_1}|^2$, which is
strictly one. Hence,
\begin{align}
  \label{eq:predictions.twophase}
  \left<\epsilon_{\B{k}_1}\,\epsilon_{\B{k}_2}\right> = \frac{(2\pi)^3}{V}\, \delta_D(\B{k}_1+\B{k}_2)\,,
\end{align}
and we see that pairing either two $\B{k}$- or $\B{q}$-modes in ${\cal E}_{\ell\ell}$ causes the respective
third wavevector to be zero, meaning that it must belong to the background. These modes do not contribute to the
correlation functions and consequently all those terms were left out in
Eq.~(\ref{eq:predictions.sixphase_expansion}).

Since one of the three Dirac delta functions is redundant and $\delta_D(\B{0}) = V/(2\pi)^3$, the Gaussian part
of Eq.~(\ref{eq:predictions.sixphase_expansion}) is of fourth order, i.e. of order $1/V^2$. The second line is a
product of two three-point correlators, which are given by Eq.~(\ref{eq:predictions.phase_average}) and by
eliminating one of the two delta functions we have,
\begin{align}
  \left<\epsilon_{\B{q}_1}\,\epsilon_{\B{k}_2}\,\epsilon_{\B{k}_3}\right>_c\, 
  \left<\epsilon_{\B{k}_1}\,\epsilon_{\B{q}_2}\,\epsilon_{\B{q}_3}\right>_c 
  = \,&\frac{(2\pi)^3}{V}\left(\frac{\sqrt{\pi}}{2}\right)^6\,\delta_D(\B{k}_1+\B{q}_1) \nonumber \\
  &\hspace{-1.5em}\times\,p^{(3)}(\B{q}_1,\B{k}_2,\B{k}_3)\,p^{(3)}(\B{k}_1,\B{q}_2,\B{q}_3)\,.
\end{align}
The quantities $p^{(N)}$ refer to the reduced $N$-th order spectra, which are defined in
Eq.~(\ref{eq:PDF.Preduced}) and have a volume dependence of $\propto V^{N/2-1}$. This implies that the
expression above is of the same order as the Gaussian term, such that higher order corrections to the
three-point correlator do not have to be taken into account. The next two contributions to ${\cal E}_{\ell\ell}$
contain connected four- and six-point correlators, which are worked out in App.~\ref{sec:four-point-phase} and
\ref{sec:six-point-phase}, giving the following results to lowest order,
\begin{align}
  \left<\epsilon_{\B{q}_1}\,\epsilon_{\B{k}_2}\,\epsilon_{\B{q}_1}\,\epsilon_{\B{q}_2}\right>_c
  \left<\epsilon_{\B{k}_3}\,\epsilon_{\B{q}_3}\right>
  = \,&\frac{(2\pi)^3}{V}\left(\frac{\sqrt{\pi}}{2}\right)^4\,\delta_D(\B{k}_3+\B{q}_3) \nonumber \\
  &\times\,p^{(4)}(\B{k}_1,\B{k}_2,\B{q}_1,\B{q}_2)\,,
\end{align}
\begin{align}
  &\left<\epsilon_{\B{k}_1}\,\epsilon_{\B{k}_2}\,\epsilon_{\B{k}_3}\,
  \epsilon_{\B{q}_1}\,\epsilon_{\B{q}_2}\,\epsilon_{\B{q}_3}\right>_c
  = \left(\frac{\sqrt{\pi}}{2}\right)^6\,
  p^{(6)}(\B{k}_1,\B{k}_2,\B{k}_3,\B{q}_1,\B{q}_2,\B{q}_3)\,.
\end{align}
Having all necessary ingredients, we can plug the phase correlators back into Eq.~(\ref{eq:predictions.covdef})
and perform the trivial Fourier integrations over the delta functions. Summing up all contributions in
Eq.~(\ref{eq:predictions.sixphase_expansion}), it is possible to show that
\begin{align}
  \label{eq:predictions.ana_covl}
  \left<\delta \hat{\ell}_i\,\delta \hat{\ell}_j\right> = \left[\frac{\left(r_i\,r_j\right)^3}{V^2}\right]^{3/2}
  \Bigg\{{\cal C}_G + {\cal C}_T + {\cal C}_{B^2} + {\cal C}_{P_6} \Bigg\}\,,
\end{align}
with 
\begin{align}
  \label{eq:predictions.ana_covG}
  {\cal C}_G \equiv\,&\underset{\substack{|\B{k}_1|,|\B{k}_2|,\\|\B{k}_1+\B{k}_2| \leq
      2\pi/r}}{\iint}\hspace{-0.25em}\frac{\D{^3\B{k}_1}}{k_f^3}\,\frac{\D{^3\B{k}_2}}{k_f^3} \nonumber \\
  &\times\,\Bigg[j_0(|2\B{k}_1+\B{k}_2|\,r_i){\cal J}(\B{k}_1,\B{k}_2,r_j) + (r_i \leftrightarrow r_j) \Bigg]\,,
\end{align}
\begin{align}
  {\cal C}_T \equiv\,&\left(\frac{\sqrt{\pi}}{2}\right)^4\underset{|\B{q}| \leq
    2\pi/r}{\int}\hspace{-0.25em}\frac{\D{^3\B{q}}}{k_f^3} \underset{\substack{|\B{k}_1|,|\B{k}_1+\B{q}|\\\leq
      2\pi/r_i}}{\int}\hspace{-0.25em}\frac{\D{^3\B{k}_1}}{k_f^3}\,
  \underset{\substack{|\B{k}_2|,|\B{k}_2+\B{q}|\\\leq
      2\pi/r_j}}{\int}\hspace{-0.25em}\frac{\D{^3\B{k}_2}}{k_f^3} \nonumber \\
  &\times\,{\cal J}(\B{q},\B{k}_1,r_i)\,{\cal J}(\B{q},\B{k}_2,r_j)\,\nonumber \\
  &\times\,p^{(4)}(\B{q},\B{k}_1,\B{k}_2,-\B{q}-\B{k}_1-\B{k}_2)\,,
\end{align}
\begin{align}
  {\cal C}_{B^2} \equiv\,&\left(\frac{\sqrt{\pi}}{2}\right)^6\underset{|\B{q}| \leq 2\pi/r}{\int}\hspace{-0.25em}
  \frac{\D{^3\B{q}}}{k_f^3}\,\Bigg[\,\Bigg( \underset{\substack{|\B{k}_1|,|\B{k}_1+\B{q}|\\ \leq
      2\pi/r_i}}{\int}\hspace{-0.25em}
  \frac{\D{^3\B{k}_1}}{k_f^3}\,{\cal J}(\B{q},\B{k}_1,r_i) \Bigg.\Bigg.\nonumber \\
  &\Bigg.\Bigg.\times\,p^{(3)}(\B{q},\B{k}_1,-\B{q}-\B{k}_1)\Bigg) \times\Bigg(r_i\leftrightarrow
  r_j\Bigg)\Bigg]\,,
\end{align}
\begin{align}
  {\cal C}_{P_6}
  \equiv\,&\left(\frac{\sqrt{\pi}}{2}\right)^6\hspace{-0.5em} 
  \underset{\substack{|\B{k}_1|,|\B{k}_2|,\\|\B{k}_1+\B{k}_2|\leq
      2\pi/r_i}}{\iint} \hspace{-0.5em}\frac{\D{^3\B{k}_1}}{k_f^3}\,\frac{\D{^3\B{k}_2}}{k_f^3}
  \hspace{-0.5em}\underset{\substack{|\B{q}_1|,|\B{q}_2|,\\|\B{q}_1+\B{q}_2| \leq 2\pi/r_j}}{\iint}
  \hspace{-0.5em}\frac{\D{^3\B{q}_1}}{k_f^3}\,\frac{\D{^3\B{q}_2}}{k_f^3} \nonumber \\
  &\times\,j_0(|\B{k}_1-\B{k}_2|\,r_i)\,j_0(|\B{q}_1-\B{q}_2|\,r_j) \nonumber \\
  &\times\,p^{(6)}(\B{k}_1,\B{k}_2,\B{k}_3,\B{q}_1,\B{q}_2,\B{q}_3)\,,
\end{align}
where we have defined the kernel function
\begin{align}
  {\cal J}(\B{k}_1,\B{k}_2,r) \equiv 2j_0(|\B{k}_1-\B{k}_2|\,r) + j_0(|2\B{k}_1+\B{k}_2|\,r)\,.
\end{align}
Furthermore, $k_f \equiv 2\pi/V^{\frac{1}{3}}$ denotes the fundamental frequency and $r \equiv
\text{max}\{r_i,\,r_j\}$. We note that Eq.~(\ref{eq:predictions.ana_covl}) closely resembles the covariance of
the bispectrum \citep[see][]{Sefusatti2006} with a Gaussian term, one that is quadratic in the bispectrum as
well as terms proportional to the trispectrum and the sixth-order spectrum. However, one important difference is
that there is no cosmology dependence in the Gaussian contribution, which is a direct consequence of the
two-point phase correlator in Eq.~(\ref{eq:predictions.twophase}), that carries no information either. That
means that Eq.~(\ref{eq:predictions.ana_covG}) is just an algebraic expression depending on the scales $r_i$ and
$r_j$, and for the variance we get explicitely,
\begin{align}
  \label{eq:predictions.ana_varlGauss}
  \text{Var}\left(\hat{\ell}(r)\right) \approx 0.25\,\frac{r^3}{V}\,.
\end{align}
%


\subsection{Cross-covariance matrix between $\hat{P}$ and $\hat{\ell}$}
\label{sec:crosscov}

Analogously to the last section we can now derive the cross-covariance between LCF and power spectrum. In this
case we deal with a mixed five-point correlator of phase factors and amplitudes,
\begin{align}
  {\cal E}_{P\ell} \equiv
  \left<\delta_{\B{k}_1}\,\delta_{\B{k}_2}\,\epsilon_{\B{q}_1}\,\epsilon_{\B{q}_2}\,\epsilon_{\B{q}_3}\right>
  -
  \left<\delta_{\B{k}_1}\,\delta_{\B{k}_2}\right>\,\left<\epsilon_{\B{q}_1}\,\epsilon_{\B{q}_2}\,\epsilon_{\B{q}_3}
  \right>\,.
\end{align}
Let us again begin by splitting this expression into its connected correlators, giving
\begin{align}
  \label{eq:predcitions.fivephase_expansion}
  {\cal E}_{P\ell} =
  &\left[\left<\delta_{\B{k}_1}\,\epsilon_{\B{q}_1}\right>\,
    \left<\delta_{\B{k}_2}\,\epsilon_{\B{q}_2}\,\epsilon_{\B{q}_3}\right>_c
    + \text{sym.}(3)\right] + \left(\B{k}_1 \leftrightarrow
  \B{k}_2\right) \nonumber \\ &+
  \left<\delta_{\B{k}_1}\,\delta_{\B{k}_2}\,\epsilon_{\B{q}_1}\,\epsilon_{\B{q}_2}\,\epsilon_{\B{q}_3}\right>_c\,,
\end{align}
where we have left out all terms that give rise to background modes as before. Due to statistical homogeneity
the mixed two-point correlator is simply the average of the amplitude $|\delta_{\B{k}}|$, which must be
evaluated using the joint PDF of Fourier modes described in App.~\ref{sec:PDF}. Assuming temporarily that we are
having a discrete set of modes, the Gaussian part is given by (see Eqs.~\ref{eq:PDF.GPDF} and
\ref{eq:PDF.interval})
\begin{align}
  \left<\delta_{\B{k}}\,\epsilon_{\B{q}}\right> &= \sqrt{V\,P(k)}\,\int \prod_{\B{p} \in \text{uhs}}\,2A_{\B{p}}
  \text{e}^{-A_{\B{p}}^2}\,\D{A_{\B{p}}}\,\frac{\D{\theta_{\B{p}}}}{2\pi}\,A_{\B{k}}\,
  \text{e}^{i\left(\theta_{\B{k}}+\theta_{\B{q}}\right)} \nonumber \\
  &= \frac{\sqrt{\pi}}{2}\sqrt{V\,P(k)}\int \prod_{\B{p} \in \text{uhs}}\frac{\D{\theta_{\B{p}}}}{2\pi}\,
  \text{e}^{i\left(\theta_{\B{k}}+\theta_{\B{q}}\right)} \nonumber \\
  &= \frac{\sqrt{\pi}}{2}\sqrt{V\,P(k)}\,\delta^K_{\B{k}+\B{q}}\,,
\end{align}
where the products run over all modes $\B{p}$ in the upper half sphere (uhs), defined by $p_z \geq 0$. In going
from the first to the second line we have made use of Eq.~(\ref{eq:PDF.int_amplitudes}) to do the integrals over
$A_{\B{p}}$ and then we see that the remaining integrals only give a non-vanishing result if the two phases
cancel out each other. The square root factor of the power spectrum enters because of our choice of
normalization, i.e. $|\delta_{\B{k}}| \equiv \sqrt{V\,P(k)}\,A_{\B{k}}$. Finally, taking the continuum limit,
\begin{align}
  \left<\delta_{\B{k}}\,\epsilon_{\B{q}}\right> = \frac{(2\pi)^3}{V}
  \frac{\sqrt{\pi}}{2}\sqrt{V\,P(k)}\,\delta_D(\B{k}+\B{q})\,.
\end{align}
In a similar manner we can compute the mixed three-point correlator and we consider the case where $\B{q}_1$,
$\B{q}_2 \in \text{uhs}$, while $\B{k} \in \text{lhs}$. The lowest order term that is contributing is
proportional to the reduced bispectrum and we have,
\begin{align}
  \left<\delta_{\B{k}}\,\epsilon_{\B{q}_1}\,\epsilon_{\B{q}_2}\right>_c = &\,\sqrt{V\,P(k)}\,\int \prod_{\B{p} \in
    \text{uhs}}\,2A_{\B{p}}\text{e}^{-A_{\B{p}}^2}\,\D{A_{\B{p}}}\,\frac{\D{\theta_{\B{p}}}}{2\pi} \nonumber \\
  &\times\,\hspace{-0.85em}\sum_{\substack{\B{u}_1,\B{u}_2,\B{u}_3\,\in\,\text{uhs} \\ \B{u}_i \neq \B{u}_j}}\hspace{-0.85em}
  A_{\B{u}_1}\,A_{\B{u}_2}\,A_{\B{u}_3}\,\cos{\left(\theta_{\B{u}_1}+\theta_{\B{u}_2}-\theta_{\B{u}_3}\right)}
  \nonumber \\ &\times\,p^{(3)}(\B{u}_1,\,\B{u}_2,\,\B{u}_3)\,A_{\B{k}}\, 
  \text{e}^{i\left(\theta_{\B{q}_1}+\theta_{\B{q}_2}-\theta_{\B{k}}\right)}\,.
\end{align}
The phase factors can only be fully cancelled in the case where $\B{k}$ equals $\B{u}_3$, such that the
integrals over $A_{\B{p}}$ produce a factor of $(\sqrt{\pi}/2)^2$. For the remaining phase integrals we need to
impose the condition $\B{k}+\B{q}_1+\B{q}_2 = 0$ and hence, the final result after taking the continuum limit is
\begin{align}
  \left<\delta_{\B{k}}\,\epsilon_{\B{q}_1}\,\epsilon_{\B{q}_2}\right>_c = &\,\frac{(2\pi)^3}{V} 
  \left(\frac{\sqrt{\pi}}{2}\right)^2\sqrt{V\,P(k)}\,p^{(3)}(\B{k},\,\B{q}_1,\,\B{q}_2)\, \nonumber \\
  &\times\,\delta_D(\B{k}+\B{q}_1+\B{q}_2)\,.
\end{align}
One can check that this result holds for any configuration of the three vectors $\B{k}$, $\B{q}_1$ and
$\B{q}_2$. The calculation of the connected five-point correlator is more cumbersome and is therefore carried
out in App.~\ref{sec:mixed-five-point}, giving the simple outcome
\begin{align}
  \left<\delta_{\B{k}_1}\,\delta_{\B{k}_2}\,\epsilon_{\B{q}_1}\,\epsilon_{\B{q}_2}\,\epsilon_{\B{q}_3}\right>_c =
  \,\,&(2\pi)^3\,\left(\frac{\sqrt{\pi}}{2}\right)^3\,\sqrt{P(k_1)\,P(k_2)} \nonumber \\
  &\times\,p^{(5)}(\B{k}_1,\,\B{k}_2,\,\B{q}_1,\,\B{q}_2,\,\B{q}_3) \nonumber \\
  &\times\,\delta_D(\B{k}_1+\B{k}_2+\B{q}_1+\B{q}_2+\B{q}_3)\,.
\end{align}
Finally, assembling all these pieces in Eq.~(\ref{eq:predcitions.fivephase_expansion}) and plugging back into
Eq.~(\ref{eq:predictions.covdef}) we obtain,
\begin{align}
  \label{eq:predictions.ana_covlP}
  \left<\delta\hat{\ell}_i\,\delta \hat{P}_j\right> =
  \frac{1}{k_f^3}\left(\frac{\sqrt{\pi}}{2}\right)^3\left(\frac{r_i^3}{V}\right)^{3/2}\,
  \Bigg[{\cal C}_{PB} + {\cal C}_{P_5}\Bigg]\,,
\end{align}
with
\begin{align}
  {\cal C}_{PB} \equiv 2 \Theta_{ij}\, P(k_j) \hspace{-0.5em}\underset{\substack{|\B{q}|,|\B{q}+\B{k}_j|\leq
      2\pi/r_i}}{\int}\hspace{-0.5em} &\D{^3q}\,j_0(|\B{k}_j-\B{q}|r_i) \nonumber \\
  &\times\,p^{(3)}(\B{k}_j,\,\B{q},\,-\B{k}_j-\B{q})\,,
\end{align}
and
\begin{align}
  {\cal C}_{P_5} \equiv\,&\frac{1}{V_P(k_j)}\int_{k_j} \D{^3p}
  \hspace{-1em}\underset{\substack{|\B{q}_1|,|\B{q}_2|,\\|\B{q}_1+\B{q}_2|\leq 
      2\pi/r_i}}{\iint} \hspace{-0.5em}\D{^3q_1}\,\D{^3q_2}\,j_0(|\B{q}_1-\B{q}_2|r_i) \nonumber \\
  &\times\,P(p)\,p^{(5)}(\B{p},\,-\B{p},\,\B{q}_1,\,\B{q}_2,\,\B{q}_3)\,.
\end{align}
Here, $\Theta_{ij}$ stands for the theta function $\Theta(1-k_j\,r_i/2\pi)$, indicating that we only get a
correlation at the order of the bispectrum if the power spectrum scale $k_j$ lies outside of the region affected
by the LCF cutoff. As for the LCF auto-covariance we notice a strong similarity to the bispectrum and power
spectrum cross-covariance given in \citet{Sefusatti2006}.

\subsection{Signal-to-noise}

In summary, we see that the LCF covariance closely resembles the covariance of the bispectrum. Its leading order
term is much simpler, though, as it carries no cosmological information -- this enters only through higher order
terms. In the large-scale limit the LCF covariance is therefore independent of redshift and shot noise. On the
other hand the lowest order contribution to the cross-covariance between $\hat{P}$ and $\hat{\ell}$ contains
cosmological information from both, the power spectrum and the bispectrum.

Based on the results from the previous section, it is instructive to compare the cumulative signal-to-noise of
the LCF with that of the power spectrum in the Gaussian approximation. To plot both as a function of the maximal
mode $k_{\text{max}}$ included, we use the correspondence $r = \pi/k$, so that we can write the LCF
signal-to-noise as follows
\begin{align}
  \label{eq:predictions.s2nl}
  \left(\frac{{\cal S}}{{\cal N}}\right)^2_{\ell} = \sum_{i=1}^{i_{\text{max}}}
  \sum_{j=1}^{i_{\text{max}}}\ell_{\text{eff}}(r_i)\,\mathsf{C}^{-1}_{ij}\,\ell_{\text{eff}}(r_j)\,,
\end{align}
where $r_i = \pi/(i\,\delta k)$ and $i_{\text{max}} = k_{\text{max}}/\delta k$. The signal
$\ell_{\text{eff}}(r)$ denotes the discrete LCF subtracted by the shot noise terms that appear in the square
bracket of Eq.~(\ref{eq:predictions.phasediscrete}) and hence takes the suppression due to $\nu_{\text{eff}}$
into account (explicit expressions are given in Eqs.~\ref{eq:measurements.leff} and
\ref{eq:measurements.lshot}).

%
\begin{figure}
  \centering
  \includegraphics{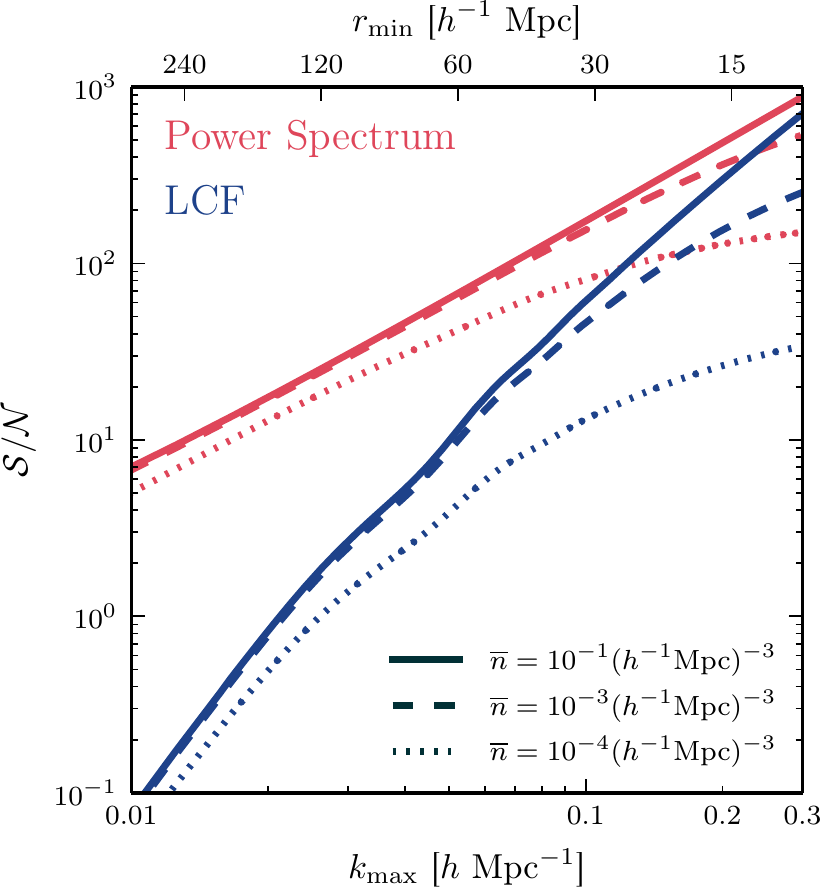}
  \caption{Cumulative signal-to-noise for the power spectrum (red) and LCF (blue) based on
    Eqs.~(\ref{eq:predictions.s2nP}) and (\ref{eq:predictions.s2nl}) as a function of the maximal mode
    $k_{\text{max}}$ or minimal scale $r_{\text{min}}$, which are related via $r_{\text{min}} =
    \pi/k_{\text{max}}$. We consider three different galaxy number densities: $\overline{n} =
    10^{-3}\,(h^{-1}\,\text{Mpc})^{-3}$ (solid lines), $10^{-4}\,(h^{-1}\,\text{Mpc})^{-3}$ (dashed)
    and $10^{-5}\,(h^{-1}\,\text{Mpc})^{-3}$ (dotted).}
  \label{fig:s2n}
\end{figure}


The Gaussian part of the covariance matrix for the power spectrum estimator in
Eq.~(\ref{eq:predictions.Pestimator}) is diagonal and given by \citep{Feldman1994}
\begin{align}
  \sigma_P^2(k) = \frac{2 (2\pi)^3}{V_P(k)\,V}\left(P(k)+\frac{1}{\overline{n}}\right)^2 = \frac{2 (2\pi)^3}{V_P(k)}\,\frac{P(k)}{V_{\text{eff}}(k)}\,,
\end{align}
with $P(k)$ meaning the discrete power spectrum subtracted by its shot noise component and $V_{\text{eff}}$ the
effective volume introduced in Eq.~(\ref{eq:predictions.Veff}). Writing $k = i\,\delta k$ the signal-to-noise is
thus
\begin{align}
  \label{eq:predictions.s2nP}
  \left(\frac{{\cal S}}{{\cal N}}\right)^2_{P} = \sum_{i=1}^{i_{\text{max}}}\frac{P(k)^2}{\sigma_P^2(k)} 
  = \sum_{i=1}^{i_{\text{max}}}\frac{V_P(k)}{2 (2\pi)^3}\,V_{\text{eff}}(k)\,.
\end{align}
In Fig.~\ref{fig:s2n} we plot Eqs.~(\ref{eq:predictions.s2nl}) and (\ref{eq:predictions.s2nP}) for various
number densities as a function of $k_{\text{max}} \,(r_{\rm min})$, where we assumed a cubical survey volume
with sidelength $L = 1.5\,h^{-1}\,\text{Gpc}$ and used a bin width of $\delta k = 2\pi/L$. As before, we adopted
a $\Lambda$CDM cosmology with parameters described in Sec.~\ref{sec:measurements} and computed the LCF in linear
PT. We see that, for most of the scales that we have considered, $k_{\rm max}<0.3\kMpc$, the
power spectrum signal-to-noise dominates over that of the LCF, being comparable only for $k_{\text{max}} \sim
0.3\,h\,\text{Mpc}^{-1}$ and the densest galaxy sample. However, the LCF signal-to-noise increases more quickly
as a function of $k_{\text{max}}$, which is a recognized feature of higher-order statistics
\citep{Sefusatti2005}. On the other hand, the LCF is also more heavily impacted by shot noise than the power
spectrum. While the difference in signal-to-noise at $k_{\text{max}} = 0.3\,h\,\text{Mpc}^{-1}$ is $\sim 10\,\%$
for the highest number density it is already $\sim 75\,\%$ for the lowest. This is because each mode
contributing to the signal-to-noise is penalized by a factor of $\nu_{\text{eff}}$, so for an $N$-th order
statistic we should expect a suppression proportional to $\nu_{\text{eff}}^N$.

Finally, we note that here we are only probing a subset of the available information in the three-point phase
correlation as the definition of the LCF in Eq.~(\ref{eq:predictions.ldef}) forces the three points of the
triangle to lie along a line -- so called degenerate triangles. Adding measurements with various other triangle
configurations is certainly going to increase the signal-to-noise. Another possibility to boost the
signal-to-noise is to find a more optimal mode cutoff than the top-hat window used in the definition of the LCF,
which could for instance be done in the Gaussian approximation of Eq.~(\ref{eq:predictions.ana_covl}).


\section{Comparison with $N$-body simulations}
\label{sec:measurements}

\subsection{Numerical simulations}
\label{sec:simulations}

In this section we present measurements of the LCF and its covariance matrix along with the cross-correlation of
the LCF with the power spectrum. These measurements are based on a set of $200$ dark matter only N-body
simulations, which were run on the \textsc{zbox}2 and \textsc{zbox}3 supercomputers at the University of Zurich
\citep[see Sec.~6 of][]{RES2009} using the \texttt{Gadget}-2 code of \citet{Springel2005}. The simulations
contain $750^3$ particles, which are enclosed in a periodic box of comoving size $L =
1500\,h^{-1}\,$Mpc. Initial conditions were set up at redshift $z=49$ based on different realizations of a
Gaussian random field and second order Lagrangian perturbation theory \citep{Crocce2006} for the displacement of
the particles. The power spectrum of the Gaussian random fields was determined from a transfer function
generated by \texttt{CMBFAST} \citep{Seljak1996} assuming a flat $\Lambda$CDM model with cosmological parameters
$\Omega_m = 0.25$, $\Omega_b = 0.04$, $\sigma_8 = 0.8$, $n_s = 1.0$ and $h = 0.7$.


\subsection{Estimating $\hat{P}$ in simulations}
\label{sec:estimation}

From the simulations we construct smooth dark matter density fields by distributing the particles onto a grid
using a cloud-in-cell (CIC) assignment scheme with $N=512$ cells per side. Each Fourier mode of the resulting
field is then corrected for the convolution with the mesh by dividing out the Fourier transform of the CIC
window function:
\begin{align}
  \delta^{d}_{\B{k}} = \frac{\delta^g_{\B{k}}}{W_{\text{CIC}}(\B{k})}\,,
\end{align}
where
\begin{align}
  W_{\text{CIC}}(\B{k}) = \prod_{i=1}^3
  \left[\frac{\sin{\left(\pi k_i/ 2k_{\text{Ny}}\right)}}{\pi k_i/ 2k_{\text{Ny}}}\right]^2\,.
\end{align}
The superscripts $d$ and $g$ denote discrete and grid quantities, respectively, and $k_{\text{Ny}} = \pi N_{\rm
  grid}/L$ is the Nyquist frequency of the mesh, with $N_{\rm grid}$ being the number of mesh-cells per
dimension.

For the power spectrum estimator presented in \Eqn{eq:predictions.Pestimator} it can be shown that for a finite
periodic volume it can be rewritten for a given scale $k$ as:
\begin{align}
  \hat{P}^d(k) = \frac{V}{N_k} \sum_{|k-q_i| \leq \Delta k/2} \left|\delta^d_{\B{q}_i}\right|^2\,,
\end{align}
where the sum extends over all modes within a shell of thickness $\Delta k$ centered around $k$, and where $N_k$
is the number of Fourier modes in each shell. This estimate still suffers from discreteness effects, and at late
times when the initial transients are small, an unbiased estimate is therefore obtained by subtracting the shot
noise term, i.e. $\hat{P} = \hat{P}^d - 1/\nbar$, where $\nbar=N/V$, with $N$ being the number of dark matter
particles. In the following we consider measurements of the power spectrum in $30$ bins from $k_{\text{min}} =
0.005$ till $k_{\text{max}} = 0.3\,h\,\text{Mpc}^{-1}$ and a bin width of $\Delta k =
0.01\,h\,\text{Mpc}^{-1}$. The power spectrum is also susceptible to aliasing effects, hence we also use FFT
grids for which $k_{\rm Ny}> 2k_{\rm max}$.


\subsection{Estimating $\hat{\ell}$ in simulations}
\label{sec:estimation}

The discretized version of the LCF estimator given by Eq.~(\ref{eq:predictions.lestimator}) can be written in
the form:
\begin{align}
{\rm E1:}\ \   \hat{\ell}^d(r) =
\left(\frac{r^3}{V}\right)^{3/2}\,\hspace{-1.75em}
\underset{\substack{|\B{k}_1|,|\B{k}_2|,\\|\B{k}_1+\B{k}_2|
      \leq 2\pi/r}}{\sum}\,\hspace{-0.75em}\overline{j_0}(|\B{k}_1-\B{k}_2|r)\,
  \epsilon^d_{\B{k}_1}\,\epsilon^d_{\B{k}_2}\,\epsilon^d_{-\B{k}_1-\B{k}_2}\,,\label{eq:est1}
\end{align}
where $\overline{j_0}(|\B{k}|r)$ denotes the spherical Bessel function averaged over the $k$-space volume
centred on the Fourier mode $\bk$. We found that this estimator is computationally expensive to estimate, at
least for the case where $r$ is probing small scales, since the 6D sum in \Eqn{eq:est1} runs over the majority
of Fourier modes -- the worst case being ${\cal O}\left(N_{\rm grid}^6\right)$ terms.

In order to accelerate the estimation we employ an implementation based on the real space phase fields. The
estimator is built around \Eqn{eq:l1}: we take the product of the $\epsilon_r(\bx)$ values at three different
points separated by scale $r$ and average these over all possible positions and orientations of the three points
in a line \citep[see also the appendix of][]{Eggemeier2015}. The new estimator can be written:
\begin{align}
  {\rm E2:}\ \   \hat{\ell}^d(r) = &\left(\frac{r^3}{V}\right)^{3/2}\,
  \frac{\Delta \varphi\,\Delta \vartheta}{16\pi}\sum_{\B{x}}\,
  \sum_{i,j=0}^n\,w_i\,w_j\,\sin{\left(j\,\Delta \vartheta\right)}\, \nonumber \\ 
  &\times\,\epsilon^d_{r}(\B{x})\,\epsilon^d_{r}\left(\B{x}+\br_{ij}\right)
  \,\epsilon^d_{r}\left(\B{x}-\br_{ij}\right)\,,
\end{align}
where the sum over $\bx$ averages over all points in the volume, and the sums over $i$ and $j$ discretize the
angular integration of the orientation of the line $\hat{\br}$ over all orientations, with $i$ labeling the
azimuthal angle and $j$ the angle with respect to the polar axis. The angular bin sizes are $\Delta \varphi =
2\pi/n$ and $\Delta \vartheta = \pi/n$. The weight factors come from the trapezoidal rule for numerical
integration and are either $w_i=1$ if $i=0$ or $n$ and 2 otherwise. The radial vector is defined by $\br_{ij}
\equiv r\B{e}_{r}(i,j)$, where the unit vector is specified by
\begin{align}
  \B{e}_{r}(i,j) \equiv \left(
  \begin{array}{c}
    \sin{\vartheta}\,\cos{\varphi} \\
    \sin{\vartheta}\,\sin{\varphi} \\
    \cos{\vartheta}
  \end{array} \right)
= \left(
  \begin{array}{c}
    \sin{(j\,\Delta \vartheta)}\,\cos{(i\,\Delta \varphi)} \\
    \sin{(j\,\Delta \vartheta)}\,\sin{(i\,\Delta \varphi)} \\
    \cos{(j\,\Delta \vartheta)}
  \end{array} \right) \ .
\end{align}
Note that owing to the fact that the phase field is smoothed on scale $r$ (see Eq.~\ref{eq:epsi}),
$\epsilon^d_r(\bx)$ has to be recomputed for each new scale for which $\ell$ is estimated. This however can be
rapidly performed by applying the cutoff in Fourier space and executing an inverse Fast Fourier Transform, i.e.
\begin{align}
  \epsilon^d_{r}(\B{x}) = \mathtt{iFFT}
  \left[\epsilon^d_{\B{k}}\,\Theta\left(1-\frac{k\,r}{2\pi}\right)\right].
\end{align}
Lastly, for the case of estimation in the $N$-body simulations, any point that gets mapped outside of the box is
placed back according to the periodic boundary conditions.

From a number of tests we have found that the numerical estimator above converges already for a moderate number
of bins, which is of the order $10$. That means E2 requires of the order $100\,N^3$ operations, independent of
the scale $r$, while E1 scales as $\sim \left(2 L/r\right)^6$ after the mode cutoff as taken into account. This
implies that E2 quickly becomes more efficient at scales $r \lesssim 10^{-1/3}\times 2L/\sqrt{N}$, i.e. $r
\lesssim 60\,h^{-1}$Mpc for $N=512$.

%
\begin{figure}
  \centering
  \includegraphics{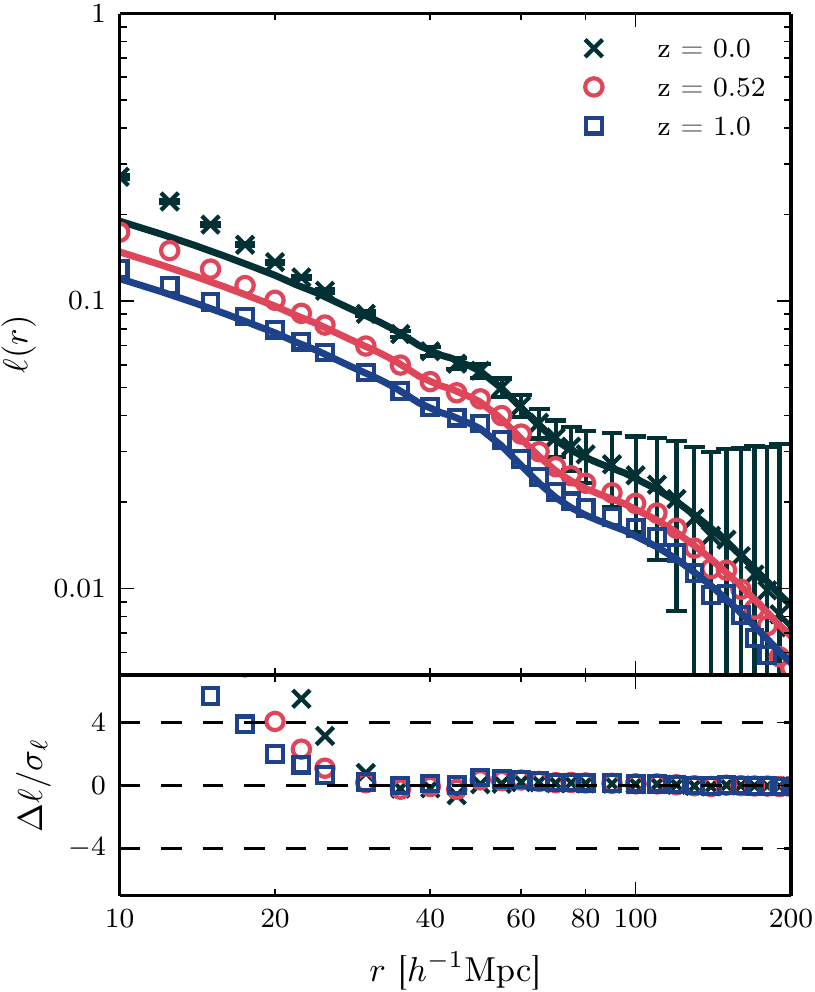}
  \caption{The top panel displays the results of the LCF measurements
    from $200$ N-body simulations (data points) for the three
    redshifts $z=0,\,0.52$ and $1$, while the $1$-$\sigma$ error bars
    for just shown for the lowest redshift sample. Solid lines of
    matching color are the corresponding predictions from tree-level
    perturbation theory. The bottom panel shows the relative
    difference between measured and predicted LCF, normalized by the
    $1$-$\sigma$ standard deviation.}
  \label{fig:linecorr_ZBOX2}
\end{figure}
%

\subsection{Comparison of estimators with simulations}

Figure~\ref{fig:linecorr_ZBOX2}, upper panel, shows the results for the matter LCF measured for three different
redshifts $z=0,\,0.52$ and $1$. The measurements were made for 30 bins with the line scale varying
$r\in[10,\,200]\,\Mpc$, which on using the relation $k\sim\pi/r$ roughly corresponds to a wavemode range of
$k\in[0.016,\,0.31]\,\kMpc$ and is hence comparable to our power spectrum measurements. The spacing of the first
$7$ bins is $2.5\,h^{-1}\,$Mpc, increasing to $5\,h^{-1}\,$Mpc for the next $11$ and the remaining $12$ bins
have a spacing of $10\,h^{-1}\,$Mpc.  The error bars show the expected variations between realisations and are
obtained from the 200 realisations. For the sake of clarity we only show the $1$-$\sigma$ error bars for the
sample with the lowest redshift. The estimates were corrected for shot noise as discussed in
\S\ref{sec:shotnoise}.

From the figure we see immediately how the LCF increases with time, which is a result of growing phase
correlations under the influence of non-linear gravity. For the same reason the LCF is a function that is mostly
decreasing with scale, as at larger $r$ the density field is in a more linear state with phases being
increasingly random. The solid lines in Fig.~\ref{fig:linecorr_ZBOX2} show the predictions from tree-level SPT
(see Eq.~\ref{eq:predictions.lPT}), and on large scales $(r>30\Mpc)$ are in good agreement with the data for the
three redshifts considered. However, on smaller scales we see that there are departures from this lowest order
prediction, which consistently underpredicts the measured LCF.

Figure~\ref{fig:linecorr_ZBOX2}, lower panel, shows the difference $\Delta \ell = \hat{\ell}-\ell$ between the
measurement and model predictions, normalized by the standard deviation. For all scales above $\sim
30\,h^{-1}\,$Mpc the difference is within the $1$-$\sigma$ interval, while for smaller scales the agreement
breaks down quickly, being already worse than $4\,\sigma$ at $r = 20\,h^{-1}\,$Mpc and the lowest
redshift. These deviations arise because non-linear corrections to the bispectrum and power spectrum become
increasingly important on these small scales. This also explains why the discrepancies are less significant for
higher redshifts. Changing the power spectrum model that enters \Eqn{eq:predictions.lPT} from the linear
spectrum to the power spectrum with corrections up to the one-loop level does not bring any improvement. On the
contrary, we note that the increase in small-scale power leads to a further suppression of the LCF and alters
the predictions by $1.4\,\sigma_{\ell}$ at $r = 40\,h^{-1}$Mpc. This suppression will be countered when using
the appropriate one-loop bispectrum, but this seems to indicate that linear theory is applicable throughout a
larger range of scales for the LCF than it is for the conventional statistical measures.

%
\begin{figure}
  \centering
  \includegraphics{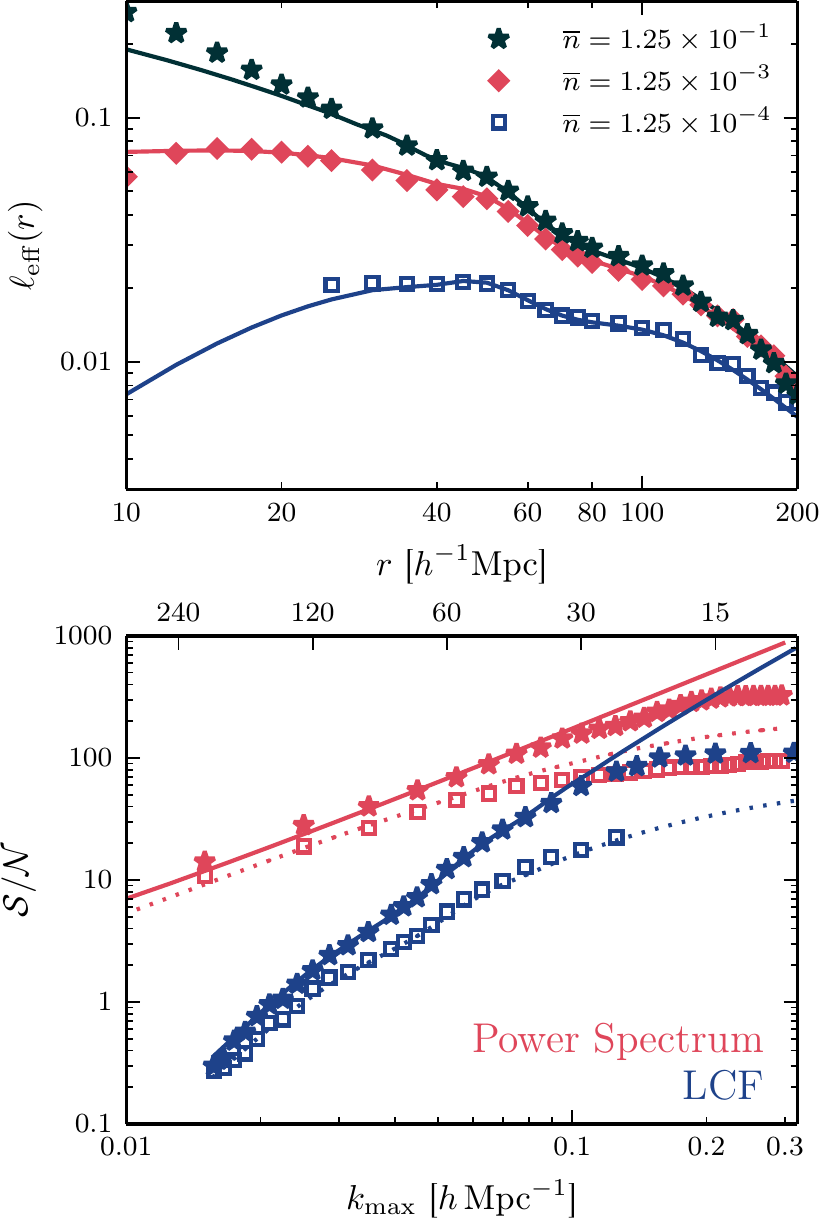}
  \caption{\textit{Upper panel:} Estimated LCF at redshift $z=0.0$ for
    different particle densities (in units of $h^{-3}\,\text{Mpc}^3$),
    corrected for additive shot noise terms. Stars mark the original
    measurements using all particles, diamonds (squares) derive from a
    subsample with $1\,\%$ ($0.1\,\%$) of the particles. Solid lines
    in the same color correspond to the tree-level
    predictions. \textit{Lower panel:} Measured cumulative
    signal-to-noise for power spectrum (red) and LCF (blue), compared
    to the approximation with Gaussian errors (solid and
    dotted lines). Symbols are the same as in the upper panel.}
  \label{fig:s2n_ZBOX2}
\end{figure}
%
%
\begin{figure}
  \centering
  \includegraphics{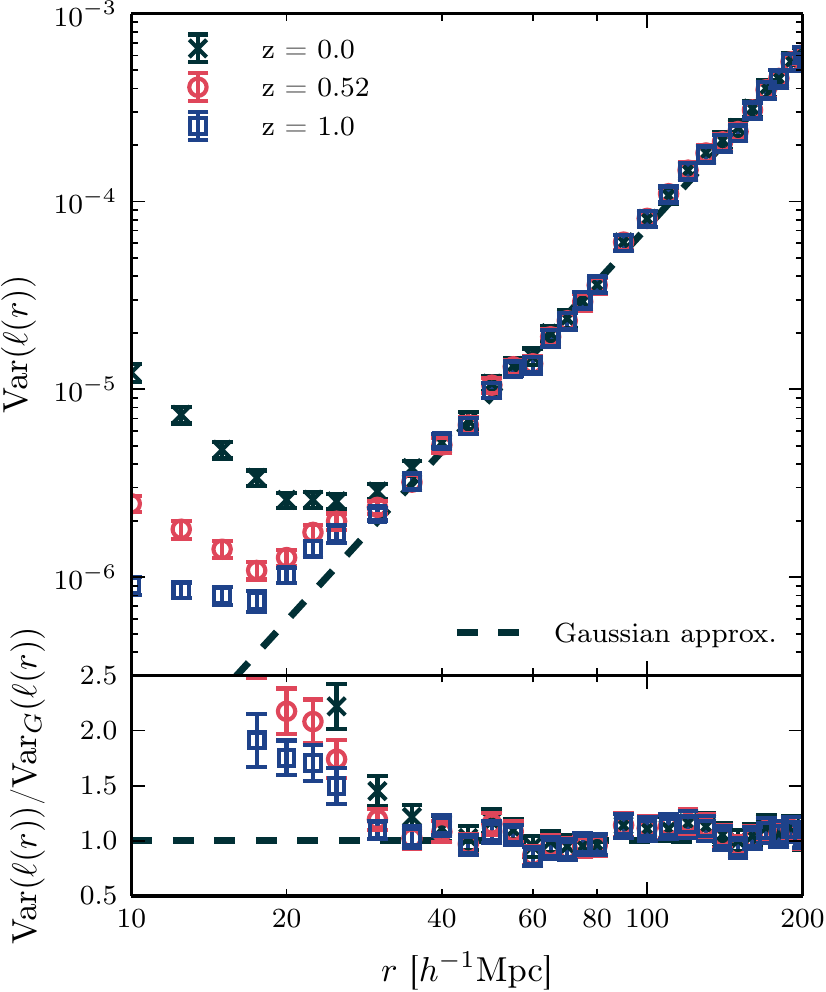}
  \caption{Variance of the LCF estimator compared to the Gaussian
    approximation $\text{Var}_G(\ell(r))$ from
    Eq.~(\ref{eq:predictions.ana_varlGauss}) for the same three
    redshifts as in Fig.~\ref{fig:linecorr_ZBOX2}. Error bars
    originate from a Jackknife resampling of the $200$ realizations.}
  \label{fig:linecorr_ZBOX2_jack}
\end{figure}
%
%
\begin{figure*}
  \centering
  \includegraphics{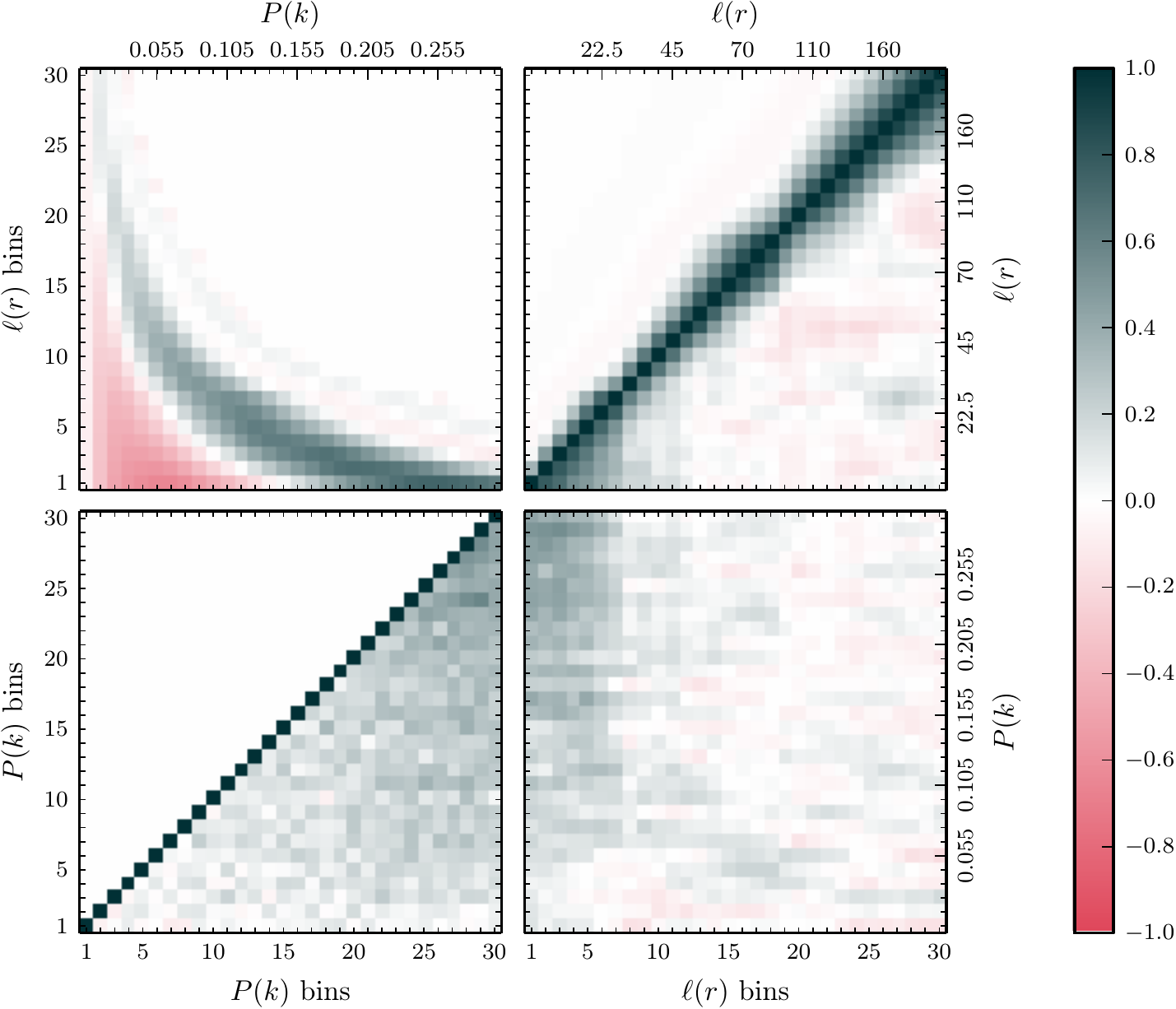}
  \caption{Correlation matrices at redshift $z=0$ with the
    auto-correlation of the LCF in the bottom left and the power
    spectrum in the top right panel. All bins below the diagonal
    derive from the measurements, all bins above are predictions based
    on the lowest order contributions to either the auto- or
    cross-covariance and linear perturbation theory. Note that on
    power spectrum axes smaller scales (higher $k$) are to the right,
    whereas for the LCF these lie on the left.}
  \label{fig:corr}
\end{figure*}
%


\subsection{Testing the effects of shot-noise on $\hat{\ell}$}

In the measurements above where all the dark matter particles have been used to construct the density field,
shot noise has no discernible effect. To test the suppression from discreteness derived in
Sec.~\ref{sec:shotnoise} we therefore carry out the same measurements, but coming from a subsample of particles
that is randomly selected before the particles are smoothed onto the grid. This procedure does not correspond to
a Poisson sampling of the underlying matter field, but it should serve as a close enough approximation
thereof. In the upper panel of Fig.~\ref{fig:s2n_ZBOX2} diamonds (squares) show the measured LCF for $1\,\%$
($0.1\,\%$) of the total number of particles, compared to the original measurements (stars), whereas the solid
lines of matching color are the theoretical predictions in tree-level SPT based on the bispectrum term in
Eq.~(\ref{eq:predictions.phasediscrete}). We clearly see how the measurements get increasingly suppressed with
decreasing number density and accurately follow the predictions, confirming our model.


\subsection{Testing the signal-to-noise of $\hat{\ell}$}

The lower panel of Fig.~\ref{fig:s2n_ZBOX2} contrasts the measured cumulative signal-to-noise with the idealized
case of Gaussian errors presented in Sec.~\ref{sec:autocov}, where the symbols are the same as in the upper
panel and colors distinguish between power spectrum and LCF. For both number densities we notice that the
measured signal-to-noise traces the predicted one very well up to a $k_{\text{max}}$ of $\sim
0.1\,h\,\text{Mpc}^{-1}$. Beyond this scale the measured signal-to-noise quickly flattens out because
higher-order corrections to the power spectrum and LCF covariance diminish the amount of available
information. As can be seen in the plot, this effect can be quite severe: at the largest $k_{\text{max}}$ the
power spectrum signal-to-noise is approximately reduced by a factor of three, while the LCF signal-to-noise even
suffers by a factor of seven.


\subsection{Estimating the covariance matrix}
\label{sec:estimated_covariances}

Apart from the means, it is also instructive to consider the covariance matrix of the power spectrum and LCF
estimators. To begin with, in Fig.~\ref{fig:linecorr_ZBOX2_jack} we compare the measured LCF variance in all
$30$ bins with the Gaussian approximation (dashed line) from Eq.~(\ref{eq:predictions.ana_varlGauss}) for the
same three redshifts as in Fig.~\ref{fig:linecorr_ZBOX2}. The error bars were estimated via Jackknife
resampling, meaning we first computed the variance $\sigma_{\ell}^2$ from the full sample of $N_{\text{real}} =
200$ realizations and subsequently from $N_{\text{real}}$ different subsamples, each giving
$(\sigma^{(i)}_{\ell})^2$, in which the $i$-th realization has been left out. The error on the variance is then
computed as follows \citep{Norberg2009}:
\vspace*{1em}
\begin{align}
  \delta \sigma_{\ell}^2 = \sqrt{\frac{N_{\text{real}}-1}{N_{\text{real}}}\sum_{i=1}^{N_{\text{real}}}
    \left[\left(\sigma^{(i)}_{\ell}\right)^2-\sigma_{\ell}^2\right]^2}\,. \\ \nonumber
\end{align}
From the figure we learn that on large scales higher-order variance terms are clearly negligible and the
measured variance displays the expected $r^3$-scaling of the Gaussian approximation
(Eq.~\ref{eq:predictions.ana_varlGauss}). However, at a scale of $\sim 30\,h^{-1}$Mpc this agreement breaks
down, the variance reaches a minimum and starts increasing with declining scales, which marks the onset of
higher-order corrections. While the Gaussian variance is independent of redshift, higher-order terms are not and
thus higher redshifts show smaller deviations from the Gaussian approximation. The scale at which the
higher-order terms become important coincides with the scale where the measured signal-to-noise was observed to
flatten out in Sec.~\ref{sec:estimation} but also with the point at which tree-level SPT breaks down
(cf. Fig.~\ref{fig:linecorr_ZBOX2}).

Figure~\ref{fig:corr} shows the full auto- and cross-correlation matrices for the power spectrum (bottom left
panel) and LCF (top right panel), where the correlation coefficient $\mathsf{r}_{ij}$ is defined to be:
\vspace*{1em}
\begin{align}
  \mathsf{r}_{ij} = \frac{\left<\delta X_i\,\delta
    X_j\right>}{\sqrt{\left<\delta X_i^2\right> \left<\delta
      X_i^2\right>}}\,. \\[-0.25em] \nonumber
\end{align}
The figure presents the measurements from simulations as all of the bins below the diagonal, whereas all bins
above denote the theoretical prediction from the respective lowest order contributions. In the case of the
prediction for the cross-correlation (top left quadrant) this results from the bispectrum term in
Eq.~(\ref{eq:predictions.ana_covlP}), computed at tree-level in SPT. Apart from some noise in the measurements
we notice that both auto-correlation matrices are very well reproduced by their Gaussian approximations on large
scales. On smaller scales, though, different bins become increasingly correlated with each other, which is
underpredicted by the lowest order contributions.

The measured cross-correlation matrix (bottom right quadrant) indicates that the power spectrum and LCF are
largely but not entirely independent of each other. The small scale LCF bins seem to be reasonably correlated
with most of the power spectrum bins. A qualitatively similar behaviour can be seen from the theoretical
computation in the top left quadrant, which displays an arc with moderate correlations for power spectrum and
LCF bins that are related by $k = \pi/r$. On large scales these correlations are of the order $\sim 0.2$ and are
therefore only hardly identifiable in the measured data, but we do recognize positive correlations along the
position of the arc. For small LCF scales and small power spectrum modes the cross-correlation is predicted to
become negative, which is not seen in the data. However, in this regime we have to expect the breakdown of
tree-level SPT as well as the influence of the next order term in Eq.~(\ref{eq:predictions.ana_covlP}).


\section{Detectability of the LCF in future surveys}
\label{sec:detection}


\subsection{Detectability of the LCF in galaxy survey data}

Before we move on to discuss the LCF's sensitivity on various cosmological parameters, we consider the
significance at which the LCF might be detected in a hypothetical galaxy survey. Taking the null hypothesis to
be the absence of any signal, the $\chi^2$ of a detection is simply given by
\begin{align}
  \chi^2_{\ell} = \sum_{i,j=1}^{N_{\text{bin}}}
  \hat{\ell}_i\,\mathsf{\hat{C}}^{-1}_{ij}\,\hat{\ell}_j\,.
\end{align}
Using all $30$ bins of our measurements we obtain $\chi^2_{\ell} \approx 10^4$, where we have accounted for the
fact that the inverse of the estimated covariance matrix is not an unbiased estimate of the inverse and applied
the Anderson--Hartlap factor \citep{Hartlap2007}, such that
\begin{align}
  \hat{\mathsf{C}}^{-1} = \frac{N_{\text{real}}-N_{\text{bin}}-2}{N_{\text{real}}-1}\,\hat{\mathsf{C}}^{-1}_{*}\,,
\end{align}
where $\hat{\mathsf{C}}^{-1}_*$ is the algebraic inverse of the measured covariance matrix. The $\chi^2$ above
is the expected value for a measurement from a single simulation box, which has a volume of $V_{\text{box}} =
(1.5\,h^{-1}\text{Gpc})^3$. As the errors scale inversely with volume, the $\chi^2_{\ell}$ for a survey of
volume $V$ and an ideal box-like geometry is thus
\begin{align}
  \label{eq:cosmoinfo.rescale}
  \chi^2_{\ell}(V) = 10^4\,\frac{V}{V_{\text{box}}}\,.
\end{align}
A $5$-$\sigma$ detection from $30$ data points corresponds to a $\chi^2\sim 85$ and this could already be
achieved by a survey with volume $V \approx 0.03\,h^{-3}\text{Gpc}^3$.


\subsection{Detectability of BAO features in the LCF}

As was noted in previous plots (e.g. see Figs~\ref{fig:lgalaxy-shot} and \ref{fig:linecorr_ZBOX2}), the LCF
signal displays slight wiggles.  These are due to BAO imprinted in the matter distribution \citep[for a review
of the physics of BAO see][]{Weinbergetal2013}. It is therefore interesting to ask: what size does our idealized
survey need to be in order to detect these features at a given confidence level?  To answer this question we
compute again the LCF in tree-level SPT, but this time for a featureless input power spectrum, obtained using
the no-wiggle fitting formula of \citet{Eisenstein1998}. The relative difference
$\left[\hat{\ell}(r)-\ell_{\text{nw}}(r)\right]/\ell_{\text{nw}}(r)$ between the measured and the no-wiggle LCF
enables us to isolate the BAO features more clearly.

\begin{figure}
  \centering
  \includegraphics{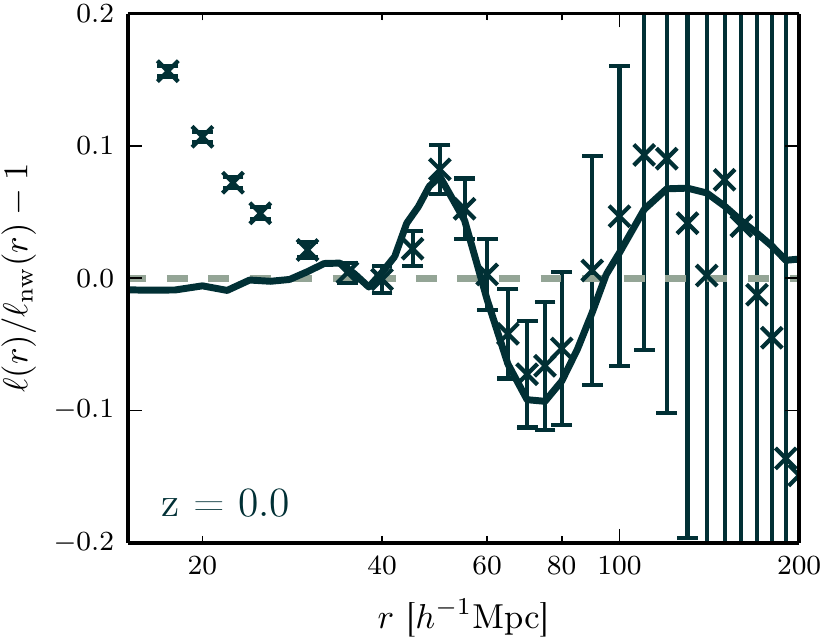}
  \caption{Relative difference between measured and no-wiggle LCF,
    obtained from tree-level perturbation theory and a featureless
    power spectrum at redshift $z=0.0$. The solid line represents the
    overall tree-level prediction. Error bars are scaled to match a
    survey of $V = 38\,h^{-3}\,\text{Gpc}^3$.}
  \label{fig:lcf_dewiggled}
\end{figure}

Figure~\ref{fig:lcf_dewiggled} shows the results of this operation for our theoretical predictions from SPT
(solid black line) and our measurements from simulations (crosses). The shape of this function can be
understood as follows, the BAO signal in the two-point function has a local maximum at roughly $r\sim100\Mpc$,
which is an imprint of the sound horizon scale at recombination. Considering the LCF, this correlates three
points along a line, each separated by distance $r$, and so there appears two values of $r$ that would produce a
resonance with the BAO scale: one when $r\sim 50\,\Mpc$ (i.e. the distance between points 3 and 1) and the
second when $r\sim 100\,\Mpc$ (i.e the separation between points 2 and 1).

Analogously to the procedure of the previous subsection, let us take the standpoint that the no-wiggle LCF
represents the null-hypothesis. Hence the $\chi^2$ for detecting the BAO features can be written:
\begin{align}
  \chi^2_{\text{BAO}} = \sum_{i,j=1}^{N_{\text{bin}}}\left[\hat{\ell}_i-\ell_{\text{nw}}(r_i)\right]
  \,\hat{\mathsf{C}}^{-1}_{ij}\,\left[\hat{\ell}_j-\ell_{\text{nw}}(r_j)\right]\,.
\end{align}
Taking all $21$ bins in the range from $40$ to $200\,h^{-1}$Mpc we get $\chi^2_{\text{BAO}} \approx
3.9$. Requiring a $3$-$\sigma$ confidence level ($\chi^2 \approx 44$ for $21$ data points), this translates into
a minimal survey volume of \mbox{$V\sim 38\,h^{-3}\text{Gpc}^3$}.  This is within reach of upcoming galaxy
surveys like DES, and certainly the Stage IV dark energy missions such as Euclid and LSST.


\section{Cosmological Information}
\label{sec:cosmoinfo}

We now turn to address the question of the cosmological information content of the LCF, where the principal aim
is to unveil which parameters or combination of parameters are best constrained by the LCF and also how it may
help to tighten existing constraints obtained from the combination of the galaxy power spectrum and a
Planck-like CMB measurement. Note that our main intention here is not to produce forecasts for a particular
survey, but simply to provide a generic assessment of the possible relative gains from measuring the LCF. Thus
in what follows we will make various simplifying assumptions. In paticular, we will neglect all effects that
lead to anisotropies in the clustering of galaxies, such as redshift space distortions and the
Alcock-Paczy\'{n}ski effect. Furthermore, we do not take any specifics of the galaxy population being surveyed
into account and simply assume an ideal box-like geometry with a constant galaxy number density
throughout. However, where possible we will try to add some validation for the choices that we make.


\subsection{Formalism and assumptions}
\label{sec:formalism_assumptions}

To begin we assume that the joint likelihood function for both the LCF and the power spectrum takes the form of
a multi-variate Guassian:
\begin{align}
  \label{eq:cosmoinfo.likelihood}
  {\cal L} = \frac{1}{\sqrt{(2\pi)^n|\B{\mathsf{C}}|}}\,\exp{\left[-\frac{1}{2}(\B{x}-\B{\mu})^T\,
    \B{\mathsf{C}}^{-1}(\B{x}-\B{\mu})\right]}\,,
\end{align}
where $\B{x}$ is the vector containing the measured data, i.e
$\bx^T=\{\hat{P}_1,\dots,\hat{P}_{m},\hat{\ell}_1,\dots,\hat{\ell_n}\}$ with mean $\B{\mu} =
\left<\B{x}\right>$, and $\B{\mathsf{C}}$ is the measured covariance matrix of dimension $(m+n)\times(m+n)$.
\citet{Takahashi2009} have shown that the probability distribution of the power spectrum estimator is indeed
very well approximated by a Gaussian distribution, over a broad range of scales.


\begin{figure}
  \centering
  \includegraphics{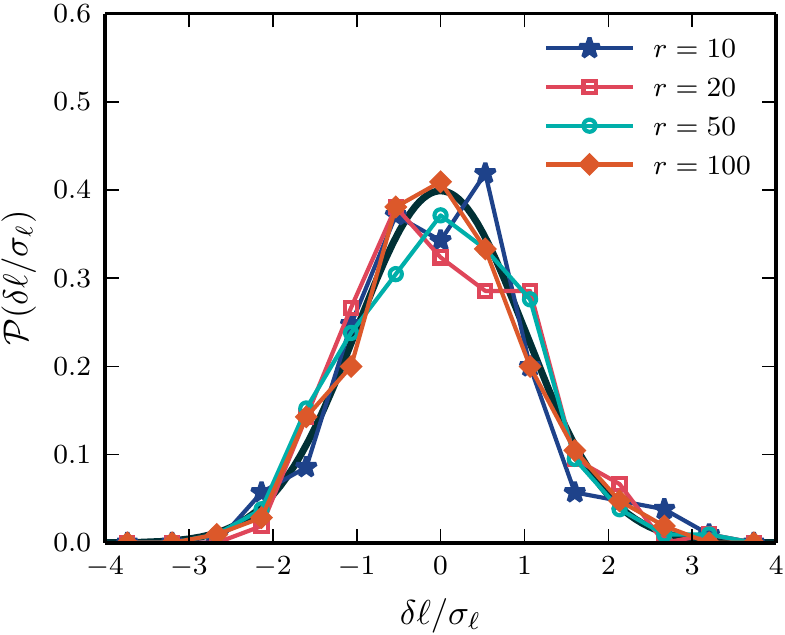}
  \caption{Probability distribution of LCF estimator at redshift
    $z=0.0$ as a function of the difference $\delta\ell =
    \hat{\ell}_i-\left<\hat{\ell}_i\right>$, normalized by the
    standard deviation. Plotted are the results for various scales
    (given in units of $h^{-1}$Mpc) and a Gaussian with zero mean and
    unit variance (black line) for reference.}
  \label{fig:probDist}
\end{figure}


For the case of the LCF there are no measurements in the literature to guide us, we therefore use our $200$
realizations to determine the LCF probability distribution at four different scales, from $10$ to
$100\,h^{-1}$Mpc. Figure~\ref{fig:probDist} shows the results form these set of measurements. The distribution
is plotted as a function of $\delta\ell = \hat{\ell}_i-\left<\hat{\ell}_i\right>$, normalized by the measured
standard deviation, such that it should approach a Gaussian distribution with zero mean and unit variance
(plotted as the thick, black line for reference). Albeit there is some scatter due to the small sample size, we
do not observe any significant indication for a strong skewness or kurtosis and thus conclude that for our
purposes here, the assumption of a Gaussian likelihood seems justified.

The parameter sensitivity of any statistic can be conveniently forecasted in the Fisher formalism. The Fisher
matrix is obtained from the logarithm of the likelihood function by taking second derivatives with respect to
the parameters of interest $\theta_i$,
\begin{align}
  \mathsf{F}_{ij} = -
  \left.\left<\frac{\partial^2 \log{{\cal L}}}{\partial\theta_i\,\partial\theta_j}\right>
  \right|_{\B{\theta}=\B{\theta}_0}\,,
\end{align}
where $\B{\theta}_0$ denotes the set of fiducial parameter values. In particular, if the likelihood is Gaussian
as in Eq.~(\ref{eq:cosmoinfo.likelihood}), it can be shown that the Fisher matrix takes the form
\citep{Tegmark1997}
\begin{align}
  \label{eq:cosmoinfo.fisherG}
  \mathsf{F}_{ij} = \frac{1}{2}\text{Tr}\left[\B{\mathsf{C}}^{-1}
    \frac{\partial\B{\mathsf{C}}}{\partial\theta_i}
    \B{\mathsf{C}}^{-1}\frac{\partial\B{\mathsf{C}}}{\partial\theta_j}\right] 
  + \frac{\partial \B{\mu}^t}{\partial \theta_i} \B{\mathsf{C}}^{-1}
  \frac{\partial \B{\mu}}{\partial \theta_j}\,.
\end{align}
By computing Eq.~(\ref{eq:cosmoinfo.fisherG}) and then taking the inverse we get the minimal achievable error on
a given parameter after marginalizing over all others,
\begin{align}
  \label{eq:cosmoinfo.margerror}
  \sigma(\theta_i) = \sqrt{\left(\mathsf{F}^{-1}\right)_{ii}}\,.
\end{align}
To facilitate the evaluation of Eq.~(\ref{eq:cosmoinfo.fisherG}) we can make a further approximation by
neglecting the first term involving derivatives of the covariance matrix. For the case of the power spectrum one
can argue that the second term on the right-hand-side of \Eqn{eq:cosmoinfo.fisherG} scales directly with the
number of Fourier modes, whereas the first term is independent and consequently is subdominant
\citep{Tegmark1997,Smith2014}. For the case of the LCF this term vanishes identically at lowest order, since, as
was shown earlier in \Eqn{eq:predictions.ana_covG}, the Gaussian part of the covariance is independent of
cosmology and as was demonstrated in Fig.~\ref{fig:linecorr_ZBOX2_jack} the Gaussian part was shown to be a
reasonable approximation for a wide range of scales.


\subsection{Parameter sensitivity}
\label{sec:paramater_sensitivity}

In the following we consider a set of nine parameters,
\begin{align}
  \B{\theta} = \left\{\Omega_m,\,\Omega_b,\,w_0,
  \,w_a,\,\sigma_8,\,n_s,\,h,\,b_1,\,b_2\right\}\,,
\end{align}
comprising the total matter and baryon densities $\Omega_m$, $\Omega_b$, the dark energy equation of state
parameters $w_0$ and $w_a$, the amplitude of density fluctuations in spheres of $8\,h^{-1}$Mpc, the scalar
spectral index $n_s$ and the dimensionless Hubble rate $h$. Lastly, we also include the two bias parameters
$b_1$ and $b_2$, using the local Lagrangian bias model introduced in Sec.~\ref{sec:galaxyLCF}. The fiducial
values that we are adopting for each of the parameters are summarized in Table \ref{tab:parameters}.
%
\begin{table}
  \centering
  \setlength{\tabcolsep}{3.5pt}
  \caption{Fiducial values of the cosmological parameters, along with
    the stepsizes $\Delta$ each parameter has been varied in either
    direction in the simulations. The bias parameters are assumed to
    be $b_1 = 1$ and $b_2 = 0$.}
  \label{tab:parameters}
  \begin{tabularx}{\columnwidth}{cccccccc}
    \hline \hline \\[-0.75em]
    \textbf{Param.} & $\Omega_m$ & $\Omega_b$ & $w_0$ & $w_a$ & $\sigma_8$ & $n_s$ & $h$ \\[0.5em] \hline \\[-0.75em]
    \textbf{Fid.} & $0.25$ & $0.040$ & $-1.0$ & $0.0$ & $0.8$ & $1.00$ & $0.70$ \\
    $\B{\Delta}$ & $\pm\,0.05$ & $\pm\,0.005$ & $\pm\,0.2$ & $\pm\,0.1$ & $\pm\,0.1$ & $\pm\,0.05$ & $\pm\,0.05$
    \\[0.5em] \hline \hline
  \end{tabularx}
\end{table}


In order to compute the Fisher matrix, we now need to determine how the power spectrum and LCF respond to
changes in these parameters, that is we need to evaluate the respective derivatives. To do that, we first
generate modified linear power spectra where one of the parameters has been changed by a step up or down
according to the values given in Table~\ref{tab:parameters}, while all others are kept at the fiducial
value. Using these spectra we compute the LCF in the tree-level approximation of Eq.~(\ref{eq:predictions.lPT})
and additionally all one-loop power spectra. That is necessary in order to include the bias parameter $b_2$ in
the power spectrum Fisher matrix because at linear order it only depends on $b_1$. The corrections due to
non-linear and non-local bias terms are summarized in App.~\ref{sec:galaxy-powerspectrum}. The derivatives are
finally obtained by taking the central finite difference of the upward and downward steps, i.e.
\begin{align}
  \label{eq:cosmoinfo.centraldiff}
  \frac{\D{X}_i(\B{\theta})}{\D{\theta}_{\alpha}} \approx \frac{X_i(\B{\theta}+\Delta\theta_{\alpha}) -
    X_i(\B{\theta}-\Delta\theta_{\alpha})}{2\,\Delta\theta_{\alpha}}\,,
\end{align}
whereas for the two bias parameters we calculate the exact derivatives from
Eqs.~(\ref{eq:predictions.lgPTterms}) and (\ref{eq:galaxy-powerspectrum.Pg}).


\begin{figure*}
  \centering
  \includegraphics[scale=0.97]{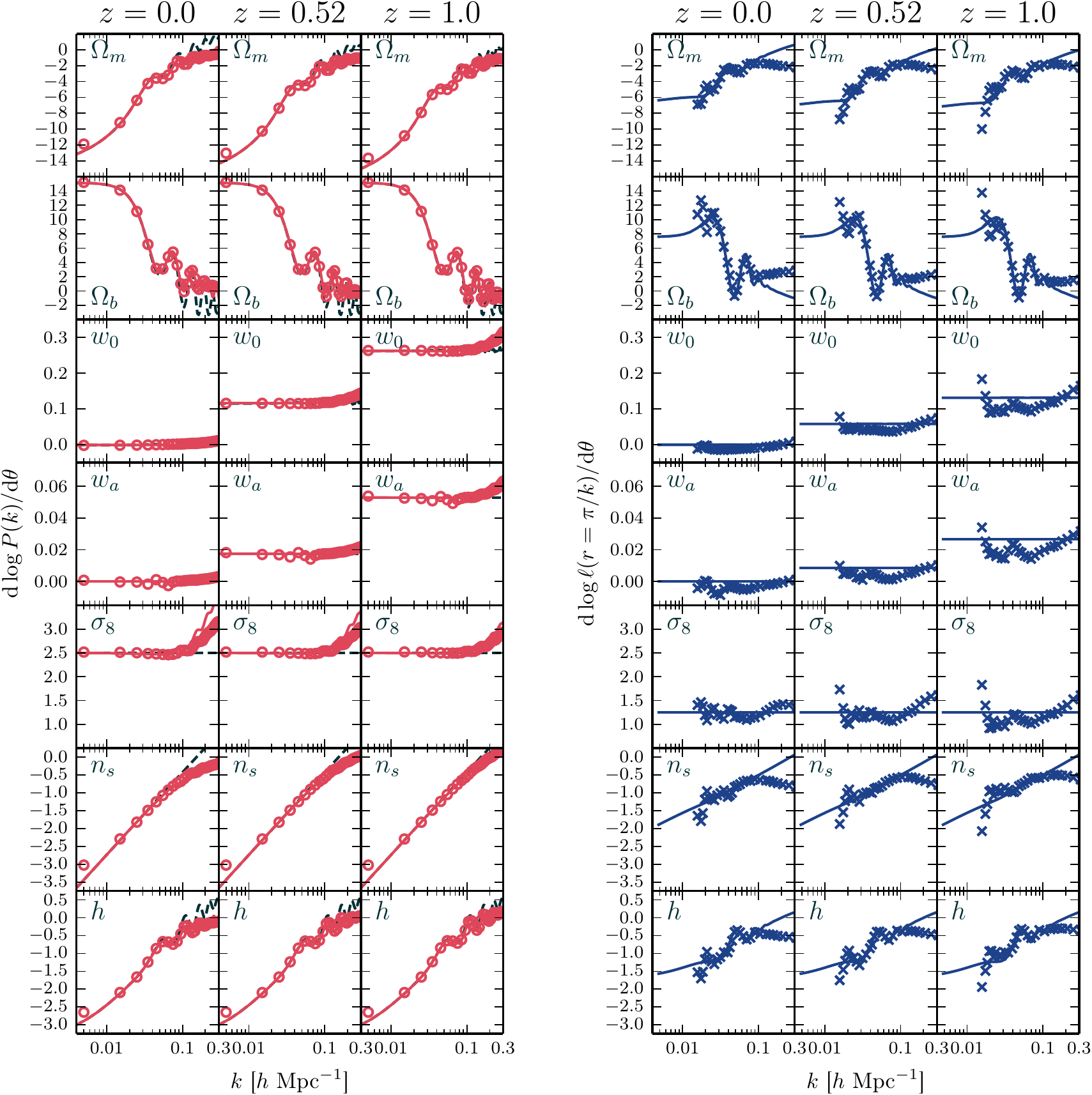}
  \caption{Time evolution of the logarithmic derivatives of power
    spectrum (left) and LCF (right) with respect to various cosmological
    parameters. Data points represent direct measurements from the
    N-body simulations described in the text, while solid lines are
    the respective model predictions. For comparison, in the left panel we also show the linear power spectrum
    derivatives as the black, dashed lines.} 
  \label{fig:par_derivative}
\end{figure*}


To check the accuracy of these model predictions, we also measure the derivatives directly from a set of
simulations whose cosmological parameters are varied in the same way as the ones given in
Table~\ref{tab:parameters} \citep[originally performed in][]{Smith2014}. The specifics of the simulations are
the same as the ones described in Sec.~\ref{sec:simulations} and for each variation as well as the fiducial
parameter set there are four realizations. To reduce the effect from sample variance, the phases of the initial
Gaussian random field of each realization are matched to the corresponding one from the fiducial
model. Derivatives are estimated as in Eq.~(\ref{eq:cosmoinfo.centraldiff}) and averaged over the four
realizations.

Figure~\ref{fig:par_derivative} shows the comparison of our predictions for the logarithmic derivatives
(depicted by solid lines) with the measurements from the $N$-body simulations (circles and crosses), and for the
various cosmological models considered.  The left panel shows the derivatives of the power spectrum and the
right the LCF. In the left panel of Fig.~\ref{fig:par_derivative} we see that the power spectrum derivatives are
reasonably well captured by the one-loop model up to the maximal scale that is being considered, $k =
0.3\,h\,\text{Mpc}^{-1}$. Note that for reference we also show the linear theory derivatives, indicated as the
dashed lines.  Considering the right panel of Fig.~\ref{fig:par_derivative} we see that the tree-level
predictions for the LCF are in reasonable agreement with the simulations up to $k = 0.1\,h\,\text{Mpc}^{-1}$,
beyond which the measured data display, in absolute terms, a larger derivative than the one predicted.


\begin{figure*}
  \centering
  \includegraphics{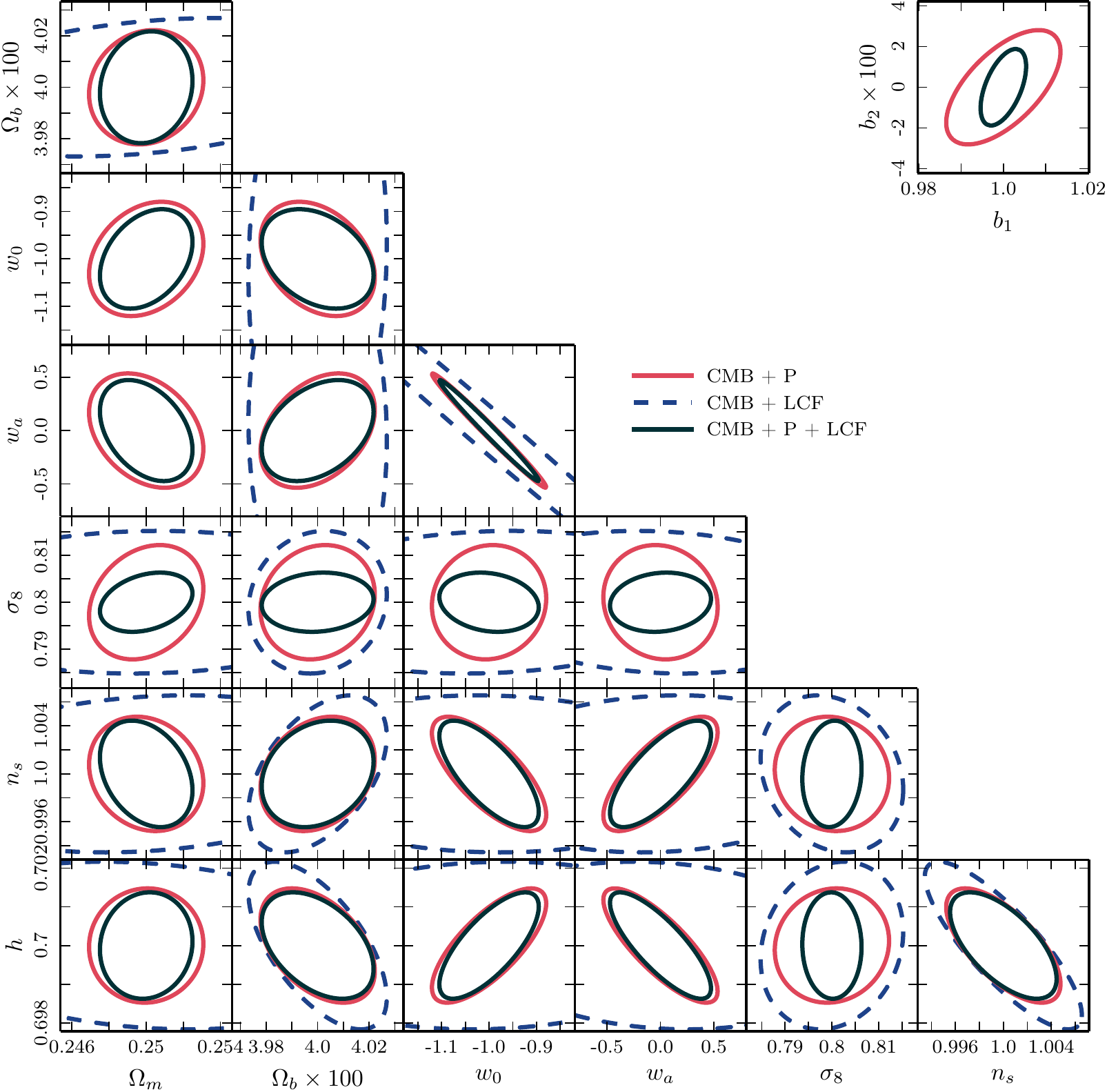}
  \caption{Forecasted $1$-$\sigma$ likelihood contours for various
    combinations of parameters, marginalized over all others. Power
    spectrum forecasts are represented by the red lines, the LCF by
    the blue dashed ones and their combination is shown in black. All
    forecasts use a cutoff scale $k_{\text{max}} =
    0.3\,h\,\text{Mpc}^{-1}$ (corresponding to $r_{\text{min}}\sim
    10\,h^{-1}$Mpc) and include information from a
    \textit{Planck}-like CMB experiment.}
  \label{fig:fisher_forecast}
\end{figure*}


On comparing the power spectrum derivatives with those of the LCF, we find that the former usually dominate over
the latter. For the three parameters $w_0$, $w_a$ and $\sigma_8$, which mainly affect the amplitude of the power
spectrum, the difference is a factor of $\sim2$ over the scales considered. This matches our expectations, since
as may be noted from inspecting \Eqn{eq:predictions.lPT} the LCF scales with the square root of the power
spectrum amplitude. On the other hand, for the remaining four parameters we find that the power spectrum
derivatives are only larger than the LCF ones on very large scales.  On smaller scales the derivatives approach
zero, signalling that the power spectrum does not provide much information on these parameters in the non-linear
regime. The LCF model also predicts nearly vanishing derivatives on small scales but the fully non-linear
measurements all saturate at some value, such that it is still possible to gain some information.


\subsection{Forecasted parameter accuracy}
\label{sec:error_ellipses}

%
\begin{table*}
  \setlength{\tabcolsep}{3pt}
  \centering
  \caption{Marginalized $1$-$\sigma$ errors for power spectrum and a
    combination of power spectrum and LCF (CMB priors are included in
    all cases). The first four columns use the cutoff scale
    $k_{\text{max}} = 0.2\,h\,\text{Mpc}^{-1}$ ($r_{\text{min}} \sim
    16\,h^{-1}$Mpc), the last four $k_{\text{max}} =
    0.3\,h\,\text{Mpc}^{-1}$ ($r_{\text{min}} \sim
    10\,h^{-1}$Mpc). Two columns for each cutoff scale correspond to
    the case where the two bias parameters have been fixed to their
    fiducial values. The percentages in the parenthesis indicate the
    improvement over the respective power spectrum results.}
  \begin{tabularx}{\textwidth}{ccccccccc}
    \hline \hline \\[-0.75em]
\label{fig:fisher_forecasts_sub}
&  P  & P + LCF & P & P + LCF & P  & P + LCF & P & P + LCF \\
    & $k_{\text{max}} = 0.2\,h\,\text{Mpc}^{-1}$ & & fixed bias & & $k_{\text{max}} = 0.3\,h\,\text{Mpc}^{-1}$ &
    & fixed bias & \\[0.5em] \hline \\[-0.75em]
    $\Delta\Omega_m$ & $0.0022$ & $0.0019\,(14\,\%)$ & $0.0016$ & $0.0015\,(4\,\%)$ & $0.0020$ & $0.0016\,(19\,\%)$ & $0.00122$ &
    $0.00116\,(5\,\%)$ \\
    $\Delta\Omega_b$ & $0.000153$ & $0.000151\,(1\,\%)$ & $0.000151$ & $0.000150\,(1\,\%)$ & $0.000146$ & $0.000143\,(2\,\%)$ &
    $0.000145$ & $0.000142\,(2\,\%)$ \\
    $\Delta w_0$ & $0.094$ & $0.084\,(11\,\%)$ & $0.088$ & $0.083\,(6\,\%)$ & $0.079$ & $0.069\,(13\,\%)$ & $0.078$ &
    $0.069\,(12\,\%)$ \\
    $\Delta w_a$ & $0.401$ & $0.370\,(8\,\%)$ & $0.388$ & $0.369\,(5\,\%)$ & $0.352$ & $0.311\,(12\,\%)$ & $0.347$ &
    $0.311\,(10\,\%)$ \\
    $\Delta\sigma_8$ & $0.0096$ & $0.0060\,(38\,\%)$ & $0.0012$ & $0.0011\,(9\,\%)$ & $0.0080$ & $0.0042\,(48\,\%)$ & $0.0010$ &
    $0.0008\,(21\,\%)$ \\
    $\Delta n_s$ & $0.0035$ & $0.0033\,(4\,\%)$ & $0.0034$ & $0.0033\,(1\,\%)$ & $0.0031$ & $0.0029\,(7\,\%)$ & $0.0030$ &
    $0.0029\,(4\,\%)$ \\
    $\Delta h$ & $0.00109$ & $0.00106\,(3\,\%)$ & $0.00108$ & $0.00106\,(2\,\%)$ & $0.00097$ & $0.00091\,(7\,\%)$ & $0.00095$
    & $0.00090\,(5\,\%)$ \\
    $\Delta b_1$ & $0.012$ & $0.005\,(60\,\%)$ & $-$ & $-$ & $0.009$ & $0.004\,(60\,\%)$ & $-$ & $-$ \\
    $\Delta b_2$ & $0.023 $& $0.020\,(13\,\%)$ & $-$ & $-$ & $0.018$ & $0.012\,(33\,\%)$ & $-$ & $-$
    \\[0.5em] \hline \hline
  \end{tabularx}
  \label{tab:errors}
\end{table*}
%

To compute the Fisher matrix we assume that our idealistic survey consists of three independent redshift slices
at $z = 0.0,\, 0.52$ and $1.0$, each of a volume $V = 3.375\,(h^{-1}\text{Gpc})^3$, so that the total Fisher
matrix based on large-scale structure is given by
\begin{align}
  \B{\mathsf{F}}^{\text{LSS}} = \B{\mathsf{F}}(z=0.0) +
  \B{\mathsf{F}}(z=0.52) + \B{\mathsf{F}}(z=1.0)\,.
\end{align}
We take the theory predictions presented in Sec.~\ref{sec:paramater_sensitivity} to model the parameter
derivatives but use the fully non-linear covariance matrices estimated from the large suite of N-body
simulations and correct the inverse for the Anderson--Hartlap factor (see Sec.~\ref{sec:detection}). When
considering combinations of power spectrum and LCF we use a bit more caution when calculating the inverse as the
two statistics have signals of widely differing orders of magnitudes. Consequently, the entries in the combined
covariance matrix will equally vary by large amounts, making the inversion process subject to numerical
errors. For that reason we first compute the correlation matrix $\B{\mathsf{r}}$, whose entries all lie in the
interval $[-1,\,1]$, and obtain its inverse via a singular value decomposition. The (uncorrected) inverse of the
covariance matrix can then be written as \citep{Smith2014}
\begin{align}
  \mathsf{C}_{*,ij}^{-1} = \frac{\mathsf{r}_{*,ij}^{-1}}{\sigma_i\,\sigma_j}\,.
\end{align}
Furthermore, we add the information coming from a CMB experiment like \textit{Planck}, acting as priors for our
parameter set. For that we initially compute the CMB Fisher matrix in a different parameter set that is more
suitable for the CMB and then transform this matrix to match our chosen large-scale structure parameters
\citep[for more details, see App. A of][]{Smith2014}. We treat the CMB information as independent from the
large-scale structure and hence the total Fisher matrix is finally given by
\begin{align}
  \B{\mathsf{F}}^{\text{tot}} = \B{\mathsf{F}}^{\text{LSS}} + \B{\mathsf{F}}^{\text{CMB}}\,.
\end{align}
Figure~\ref{fig:fisher_forecast} shows the $1$-$\sigma$ likelihood contours derived from this Fisher matrix for
various combinations of parameters, after marginalizing over all others, and a maximal mode $k_{\text{max}} =
0.3\,h\,\text{Mpc}^{-1}$ ($r_{\text{min}} \sim 10\,h^{-1}$Mpc). The error ellipses are constructed by inverting
$\B{\mathsf{F}}^{\text{tot}}$ and reducing it to a $2 \times 2$ submatrix of the desired parameters. This
submatrix is inverted back again and we determine its eigenvalues and eigenvectors, which are used as input to
plot the corresponding error ellipses. In each panel the red lines represent the case where
$\B{\mathsf{F}}^{\text{LSS}}$ is evaluated for the power spectrum alone, the blue dashed lines are for the LCF
and the black ones the combination of both measures. Note that the CMB prior is always added.

The figure illustrates that there are some substantial gains over the power spectrum plus CMB alone to be
made. In particular, the largest gains are obtained for the $\sigma_8$ parameter, which is mainly a proxy for
the amplitude of fluctuations, and the two bias parameters $b_1$ and $b_2$. There is also a more modest
improvement in the constraints on the matter density parameter $\Omega_{\rm m}$. However, for the other
parameters $\{\Omega_b,\,n_s,\,h,w_0,w_a\}$ the gains are marginal. For the first three parameters, this is not
too surprising since they are already well constrained by the CMB. The dark energy equation of state parameters
do not display a significant improvement, either. Perhaps this owes to the fact that the LCF is only very weakly
dependent on growth history and the nonlinear interaction kernel seems to be somewhat cosmology independent.


\begin{figure}
  \centering
  \includegraphics{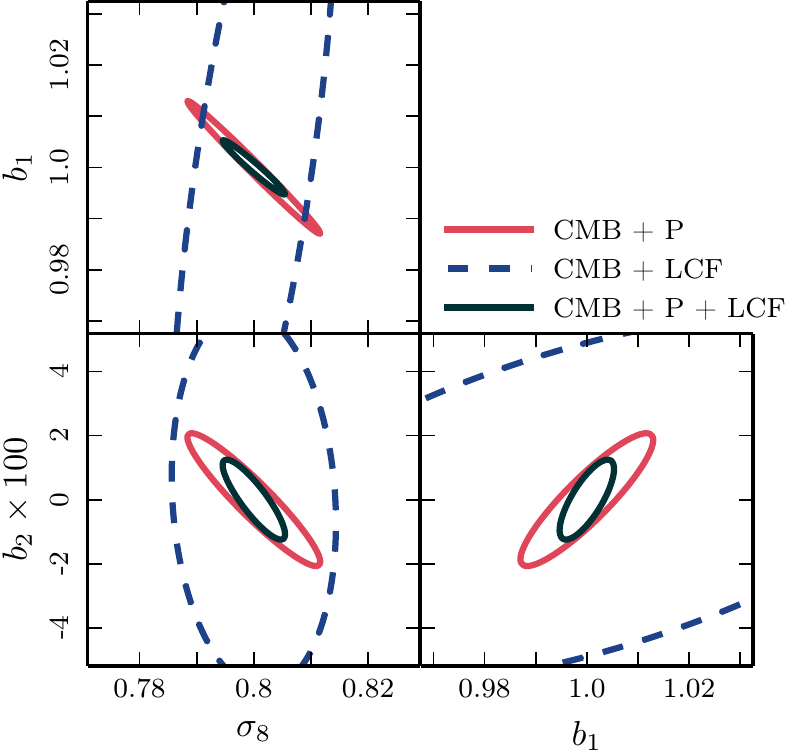}
  \caption{Forecasted $1$-$\sigma$ likelihood contours as in
    Fig.~\ref{fig:fisher_forecast} but for a subset of parameters
    including $\sigma_8$, $b_1$ and $b_2$. The cutoff scale is chosen
    to be $k_{\text{max}} = 0.3\,h\,\text{Mpc}^{-1}$.}
  \label{fig:fisher_forecast_sub}
\end{figure}


The above qualitative findings are shown more quantitatively in Table \ref{tab:errors}, which summarizes the
marginalized $1$-$\sigma$ errors for all parameters and for two different cutoff scales: the one used for
Fig.~\ref{fig:fisher_forecast} as well as the more conservative choice $k_{\text{max}} =
0.2\,h\,\text{Mpc}^{-1}$. The combinations of power spectrum and LCF (including the CMB priors) are to be found
in the sixth and second data column, respectively, where the percentages in parenthesis give the improvement
compared to the power spectrum alone and we read off that the errors for $\sigma_8$ and $b_1$ decrease by $48$
and even $60\,\%$ when LCF information is included. Interestingly, we obtain comparable improvement factors when
the lower cutoff is being used, only $b_2$ displays a larger change from $13$ to $33\,\%$, indicating that the
information coming from smaller LCF scales are particularly helpful in constraining non-linear bias. However, we
note that the improvement factors are slightly underestimated because as we have seen in
Sec.~\ref{sec:paramater_sensitivity}, our LCF model somewhat underpredicts the parameter sensitivity in the
non-linear regime.

We also consider the case where we assume that the bias parameters are known and fixed to their fiducial values,
meaning that we simply strike out all the corresponding rows and columns in $\B{\mathsf{F}}^{\text{tot}}$. The
resulting errors are given in the last two columns for each cutoff scale and we now observe a reduction of all
improvement factors, which is particularly evident for $\Omega_m$ and $\sigma_8$. That implies that for these
parameters the gain from the LCF is mainly due to a better constraint of galaxy bias. We also see that the
inclusion of a larger amount of the small scale modes now brings about slightly more significant improvements.

To further investigate where the main constraining power of the LCF is coming from, we analyze a subset of our
parameters that just comprises $\sigma_8$ and the bias parameters. The resulting likelihood contours of an
analogue Fisher matrix computation are shown in Fig.~\ref{fig:fisher_forecast_sub} and they clarify why we
obtain the comparably good improvements noted above. Even though the LCF cannot put a tight constraint on $b_1$,
it does produce small error bars for $\sigma_8$ (cf. $\Delta \sigma_8$ in the last column of Table
\ref{tab:errors}). As the degeneracy direction in the $\sigma_8$-$b_1$ plane (see top left panel in
Fig.~\ref{fig:fisher_forecast_sub}) for the LCF is fundamentally different from that of the power spectrum, a
combination of both measures lead to good constraints, which in turn carry over to $b_2$. This behaviour is
reasonable because as was shown in Sec.~\ref{sec:galaxyLCF} the LCF is independent of linear bias at lowest
order (but not of $\sigma_8$) and thus breaks the degeneracy between both parameters when combined with e.g. the
power spectrum.


\section{Discussion and Conclusions}
\label{sec:discussion}

In this paper, we have studied the ability of the LCF to constrain the cosmological model in combination with
power spectra measurements of the large-scale galaxy and CMB fluctuations.

In order to achieve this it was necessary to extend the LCF from describing matter fluctuations to those of
galaxy fluctuations. In \S\ref{sec:predictions} we did this by computing the LCF in the Lagrangian biasing
scheme. While the LCF is independent of bias in the regime where the relation between the galaxy and dark matter
overdensities is linear, we have seen that non-linearity and non-locality introduce additional terms. However,
if bias is assumed to be local in Lagrangian space \citep[this approach was adopted in recent bispectrum
measurements from BOSS by][]{Gil-Marin2016}, the LCF still breaks the degeneracy between the amplitude of
density fluctuations and the two remaining bias parameters. In comparison, this is not possible if one considers
the bispectrum alone.

We also determined the effect of shot noise on the LCF, finding that after additive contributions are removed
the signal becomes increasingly suppressed with decreasing number densities. Unlike more conventional clustering
measures, the LCF cannot be completely cleaned from shot noise and the galaxy number density must hence be
incorporated into the modelling.

In \S\ref{sec:LCFcovariance} we provided the first ever derivation of the LCF auto-covariance and its
cross-covariance with the power spectrum. We noted that there was a structural similarity for the joint
covariance we computed and that associated with the joint covariance between the power spectrum and the
bispectrum. More importantly, though, we were able to prove that, in the Gaussian limit, the lowest order
contribution to the LCF covariance was independent of cosmological parameters. For that reason it is not subject
to non-linear evolution and can therefore be predicted exactly. This property might prove to be advantageous
compared to the bispectrum, as it might allow us to produce accurate covariance matrices from relatively cheap
small-scale $N$-body simulations, matched with the analytic results in the Gaussian limit. Of course there are
additional higher order corrections which involve the trispectrum, quadratic powers of the bispectrum, and the
6-point spectrum of the phase field that could complicate this possibility.

\begin{figure}
  \centering
  \includegraphics{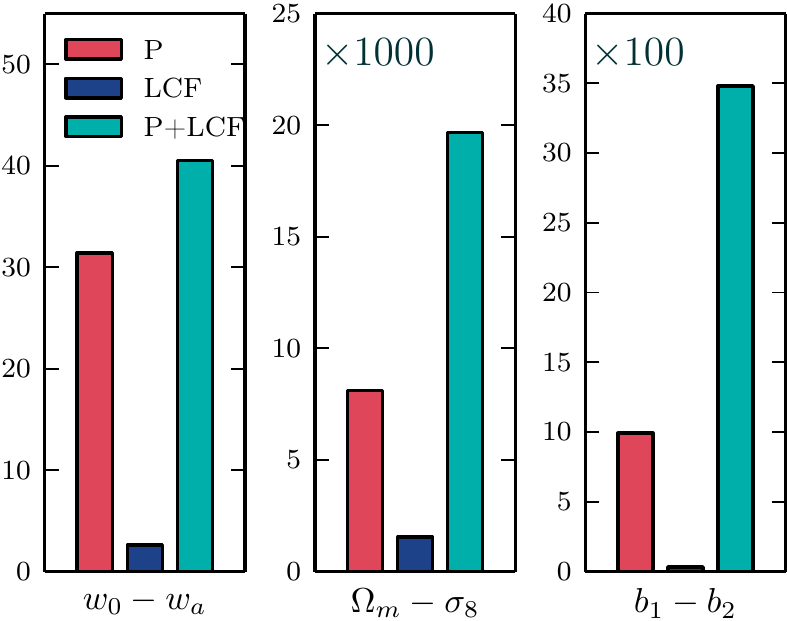}
  \caption{Figure of merit for three different parameter combinations. The bar on the left in each panel is the
    power spectrum, the middle bar marks the LCF and the last one the combination of both. All three include
    the CMB priors.}
  \label{fig:fom}
\end{figure}

In \S\ref{sec:measurements} we confronted our analytic results with measurements from a large ensemble of
$N$-body simulations, comprising a total combined volume of $675\,h^{-3}\,\text{Gpc}^3$. The simulations enabled
us to produce the first ever measurement of the LCF at the BAO scale, with enough volume, to unambiguously
detect the acoustic oscillation features. Through the simulation to simulation variance, we were also able to
produce the first ever measurement of the LCF covariance matrix and its cross-covariance with the power
spectrum. We found that the Gaussian approximation of the covariance holds down to scales $\sim 40\,h^{-1}$Mpc,
slightly above the scale where the tree-level SPT result for the LCF breaks down. As expected, the power
spectrum and LCF are mostly uncorrelated but we do detect moderate correlations of the order $\lesssim 0.4$ for
small LCF bins as well as bins of equivalent scales, i.e. those, which are related via $k \approx \pi/r$.

In addition, we also computed the signal-to-noise ratio of the LCF, from which we discovered that the power
spectrum dominates over the LCF up to $k_{\text{max}} = 0.3\,h\,\text{Mpc}^{-1}$ ($\sim
10\,h^{-1}$Mpc). However, the LCF signal-to-noise shows a stronger increase with decreasing scale. This suggests
that there is potentially more information to be gained in the LCF than the power spectrum by pushing to smaller
scales.

In \S\ref{sec:detection} we turned to assessing the detectability of the LCF in galaxy surveys. We found that
the LCF could be measured at $>5\sigma$ significance in a survey of volume $V>0.03\,h^{-3}\text{Gpc}^3$ -- this
paves the way for a very high significance measurement in modern surveys like the SDSS main sample and BOSS. On
the other hand, we found that in order to detect the BAO signature one would require a survey that spans a
volume of roughly \mbox{$V\sim 40\,h^{-3}\text{Gpc}^3$} and hence this would only likely be possible with future
Stage IV missions like Euclid and LSST.

In \S\ref{sec:cosmoinfo} we explored the main question of this paper, which was the cosmological information
content of the LCF. We considered a large-scale structure survey of total volume $\sim 10\Gpccube$, up to
$z\sim1$ -- thus comparable with Stage-III like spectroscopic missions \citep{DETF2006}. We found that when
estimates of the LCF are combined with estimates of the galaxy power spectrum and a Planck-like CMB experiment,
significant improvements may be found in constraints on the amplitude of density fluctuations $\sigma_8$
(roughly a factor of $\sim$2) and to more modest improvements in the matter density parameter $\Omega_{\rm m}$
($\sim20$\%). In addition, one is able to significantly improve constraints on the nonlinear galaxy bias
parameters $b_1$ and $b_2$ (factor $\sim2$). Expressed in terms of the figure of merit (see Fig.~\ref{fig:fom})
we obtain improvement factors of $\sim3.5$ and $\sim2.4$ for the parameter combinations $b_1$--$b_2$ and
$\Omega_{\rm m}$--$\sigma_8$, respectively\footnote{The figure of merit is defined as the inverse area enclosed
  by the $2\,\sigma$ error ellipse for any combination of two parameters \citep{DETF2006}, i.e.
  \begin{align}
    \text{FOM}(\theta_i,\,\theta_j) \equiv
    \frac{1}{\pi\sqrt{6.17\,\text{Det}\left[\B{\mathsf{F}}^{-1}(\theta_i,\,\theta_j)\right]}}\,, \nonumber
  \end{align}
  where $\B{\mathsf{F}}(\theta_i,\,\theta_j)$ is the $2 \times 2$ Fisher matrix of the parameters $\theta_i$ and
  $\theta_j$.}. On the other-hand we find no significant improvement to be gained in the traditional dark energy
figure of merit (it changes by a factor of $\sim1.3$).

As we have only used $30$ bins for our forecasts, these results suggest that the LCF provides an efficient
compression of higher-order information. This could prove to be advantageous over the bispectrum, as it
simplifies the task of generating accurate covariance matrices. Clearly, a more detailed study comparing various
higher-order statistics is necessary to make a more definitive statement and will be presented in a forthcoming
paper. Besides, by definition the LCF probes a very particular configuration of three points, so it would be
interesting to explore whether there are more optimal shapes for constraining parameters.

Finally, to facilitate the Fisher analysis in this paper, we have made a number of simplifications. In
particular, we have neglected effects from redshift space distortions; the imprint of the finite survey geometry
and survey mask on the phase field. Both of these will need to be explored in future work to arrive at more
realistic forecasts and methodology for survey analysis. Accounting for redshift space distortions should
provide additional information. Moreover, owing to the fact that in a power spectrum analysis the growth rate of
structure $f(\Omega)$ is strongly degenerate with $\sigma_8$ and $b_1$, we expect the LCF will also prove to be
effective in breaking these degeneracies and so {\em should} provide improvements in the dark energy figure of
merit. Furthermore, we have assumed throughout that the primordial density field was Gaussian. If this was not
the case, the LCF would acquire a further contribution, whose amplitude $f_{\text{NL}}$ would be naturally
degenerate with $b_1$, which means it can be potentially well constrained by a combination of power spectrum and
LCF.


\section*{Acknowledgments}

We would like to thank Joyce Byun, Donough Regan and David Seery for
useful discussions. AE acknowledges support from the STFC for his PhD
studentship funding.



\bibliographystyle{mnras}
\bibliography{complete_manuscript}





\appendix

\section{The Joint Probability Density Function of Fourier Modes}
\label{sec:PDF}

In this appendix we present results from \citet{Matsubara2003, Matsubara2006}, which will be used for the
derivation of the LCF covariance in Sec.~\ref{sec:LCFcovariance}.

Given a density field enclosed in a box of volume $V$ with discrete Fourier modes $\delta_{\B{k}}$, all of its
statistical properties are encoded in the probability density function (PDF) ${\cal P}$. Normalizing the density
modes by volume and their power spectrum we get the dimensionless variables $\alpha_{\B{k}} \equiv
\delta_{\B{k}}/\sqrt{V\,P(k)}$, in terms of which the PDF can be written as
\begin{align}
  \label{eq:PDF.PDF}
  {\cal P}[\alpha_{\B{k}}] =\,&\exp{\Bigg[\sum_{N=3}^{\infty}\frac{(-1)^N}{N!}\,
    \sum_{\B{k}_1} \cdots \sum_{\B{k}_N}\left<\alpha_{\B{k}_1}\,\cdots\,\alpha_{\B{k}_N}\right>_c} \nonumber \\
  &\times\,\frac{\partial}{\partial \alpha_{\B{k}_1}}\,\cdots\frac{\partial}{\partial \alpha_{\B{k}_N}}\Bigg]\,
  {\cal P}_G[\alpha_{\B{k}}]\,,
\end{align}
where 
\begin{align} 
  \label{eq:PDF.GPDF}
  {\cal P}_G[\alpha_{\B{k}}] = \frac{1}{2\pi} \exp{\left(-\frac{1}{2}
      \sum_{\B{k}}\alpha_{\B{k}}\alpha_{-\B{k}}\right)}
\end{align}
denotes the Gaussian PDF. The $\left<\alpha_{\B{k}_1}\,\cdots\,\alpha_{\B{k}_N}\right>_c$ refer to the $N$-th
order cumulants,
\begin{align}
  \left<\alpha_{\B{k}_1}\,\cdots\,\alpha_{\B{k}_N}\right>_c = p^{(N)}(\B{k}_1,\,\ldots\,\B{k}_N)\,
  \delta^K_{\B{k}_1+\cdots +\B{k}_N}\,,
\end{align}
which are related to normalized (and dimensionless) versions of the ordinary $N$-th order spectra $P^{(N)}$,
defined as
\begin{align}
  \label{eq:PDF.Preduced}
  p^{(N)}(\B{k}_1,\,\ldots\,\B{k}_N) \equiv
  V^{1-\frac{N}{2}}\,\frac{P^{(N)}(\B{k}_1,\,\ldots\,\B{k}_N)}{\sqrt{P(\B{k}_1)\,\cdots\,P(\B{k}_N)}}\,.
\end{align}
For mildly non-Gaussian fields we can expand the exponential in Eq.~(\ref{eq:PDF.PDF}), 
\begin{align}
  \label{eq:PDF.expansion}
  {\cal P}[\alpha_{\B{k}}] = \Bigg[1 + \sum_{n=1}^{\infty} {\cal Q}^{(n)}\Bigg]\, {\cal
    P}_G[\alpha_{\B{k}}]\,,
\end{align}
where the terms ${\cal Q}^{(n)}$ contain spectra of various orders and can be expressed as follows
\begin{align}
  \label{eq:PDF.Qn}
  {\cal Q}^{(n)} = &\sum_{m=1}^{\infty} \frac{1}{m!} \sum_{\substack{n_1,\ldots,n_m\geq 1 \\ n_1+\cdots+n_m=n}}
  \frac{1}{(n_1+2)!\cdots(n_m+2)!} \nonumber \\
  &\times\,\sum_{\B{k}_1^{(1)},\,\ldots,\,\B{k}_{n_1+2}^{(1)}} \cdots
  \sum_{\B{k}_1^{(m)},\,\ldots,\,\B{k}_{n_m+2}^{(m)}} p^{(n_1+2)}\cdots p^{(n_m+2)}
  \nonumber \\ &\times\, 
  H_{\B{k}_1^{(1)}\cdots\,\B{k}_{n_1+2}^{(1)}\cdots\,\B{k}_1^{(m)}\cdots\,\B{k}_{n_m+2}^{(m)}}\,,
\end{align}
with $H$ standing for a generalization of the Hermite polynomials,
\begin{align}
  H_{\B{k}_1\cdots\,\B{k}_n} = \frac{(-1)^n}{{\cal
      P}_G[\alpha_{\B{k}}]}\,\frac{\partial}{\partial\,\alpha_{\B{k}_1}} \cdots
  \frac{\partial}{\partial\,\alpha_{\B{k}_N}} {\cal P}_G[\alpha_{\B{k}}]\,.
\end{align}
In our Universe structure formation roughly follows the hierarchical model, meaning we have that $P^{(N)} \sim
{\cal O}\left[P(k)^{N-1}\right]$, such that any given term in the series expansion above is of the order
\begin{align}
  {\cal Q}^{(n)} \sim \varepsilon^{N-2}\,, \hspace{1.5em} \varepsilon \equiv \sqrt{\frac{P(k)}{V}}\,.
\end{align}
It follows that the expansion is only meaningful as long as this parameter $\varepsilon$ remains
small. Furthermore, this allows us to conveniently keep track of the order of each term by counting powers of
$1/\sqrt{V}$.

As we are going to evaluate ensemble averages comprising amplitudes and/or phases of Fourier modes, it is useful
to split our variables accordingly and write them as $\alpha_{\B{k}} = A_{\B{k}}\,\text{e}^{i
  \theta_{\B{k}}}$. However, due to the reality constraint, $\alpha_{\B{k}}$ and $\alpha_{\B{k}}^* =
\alpha_{-\B{k}}$ are not entirely independent from each other, which is why we restrict all summations over
wavevectors to the upper half sphere (uhs), defined by $k_z \geq 0$. In this subspace the probability for a set
of modes to take values within an infinitesimal interval is thus given by
\begin{align}
  \label{eq:PDF.interval}
  &{\cal P}(\{\alpha_{\B{k}},\,\alpha_{\B{k}}^*\})\,\prod_{\B{k}\,\in\,\text{uhs}}
  \D{\alpha_{\B{k}}} \D{\alpha_{\B{k}}^*} \nonumber \\  &= {\cal P}(\{A_{\B{k}},\,\theta_{\B{k}}\})\,\prod_{\B{k}\,\in
    \,\text{uhs}} 2 A_{\B{k}}\,\D{A_{\B{k}}}\,\D{\theta_{\B{k}}}\,,
\end{align}
where the factor $2 A_{\B{k}}$ comes from the Jacobian of the transformation. Expressing the ${\cal Q}^{(n)}$ in
terms of $A_{\B{k}}$ and $\theta_{\B{k}}$, all resulting terms can be rearranged to display a similar structure:
\begin{align}
  \label{eq:PDF.generalterm}
  \sum_{\substack{\B{k}_1,\,\ldots\,\in\,\text{uhs} \\ \B{k}_i \neq \B{k}_j}} &A_{\B{k_1}}\,A_{\B{k_2}} \cdots \cos{\left(
      \theta_{\B{k}_1} \pm \theta_{\B{k}_2} \pm \cdots\right)} \nonumber \\
  &\hspace{-0.75em}\times\,
  p^{(n_1)}(\B{k}_1,\,\ldots,\,\B{k}_{n_1})\,p^{(n_2)}(\B{k}_{n_1+1},\,\ldots,\,\B{k}_{n_1+n_2}) \cdots\,, 
\end{align}
and it is important to note that when integrating over the phases, we always get a vanishing result unless the
cosine-term is cancelled by some means. That is only possible if we correlate a number of phase factors, which
exactly matches the number of phases appearing in Eq.~(\ref{eq:PDF.generalterm}). Since all other terms in
Eq.~(\ref{eq:PDF.Qn}) give no contribution, this drastically simplifies our task of computing any particular
phase correlator. From this observation also follows that any even (odd) phase correlator can only get
contributions from even (odd) $N$-th order spectra. We will now consider some special cases, which occur in the
main part of this work.

\subsection{Four-point phase correlator}
\label{sec:four-point-phase}

We only take into account terms of the order $1/V$, which corresponds to $n=2$ in the PDF expansion of
Eq.~(\ref{eq:PDF.expansion}). After splitting each summation into separate sums over mutually different modes,
we obtain the following two terms with four different phase factors,
\begin{align}
  \label{eq:PDF.4pQ1}
  {\cal Q}^{(2)}_1 =\,&\frac{1}{3} \sum_{\substack{\B{k}_1,\,\B{k}_2,\,\B{k}_3 \\ \B{k}_i \neq \B{k}_j}}^{\text{uhs}}\hspace{-0.25em}
  A_{\B{k}_1}\,A_{\B{k}_2}\,A_{\B{k}_3}\,A_{\B{k}_{123}} \nonumber \\
  &\hspace{-2.5em}\times\,\cos{\left(\theta_{\B{k}_1}+\theta_{\B{k}_2}+\theta_{\B{k}_3}-\theta_{\B{k}_{123}}\right)}\,
  p^{(4)}(\B{k}_1,\,\B{k}_2,\,\B{k}_3,\,-\B{k}_{123})\,,
\end{align}
\begin{align}
  \label{eq:PDF.4pQ2}
  {\cal Q}^{(2)}_2 =\,&\sum_{\substack{\B{k}_1,\,\B{k}_2,\,\B{k}_3 \\ \B{k}_i \neq \B{k}_j}}^{\text{uhs}}\hspace{-0.25em}
  A_{\B{k}_1}\,A_{\B{k}_2}\,A_{\B{k}_3}\,A_{\B{k}_{123}} \nonumber \\
  &\hspace{-2.5em}\times\,\cos{\left(\theta_{\B{k}_1}+\theta_{\B{k}_2}-\theta_{\B{k}_3}-\theta_{\B{k}_{123}}\right)}\,
  p^{(4)}(\B{k}_1,\,\B{k}_2,\,-\B{k}_3,\,-\B{k}_{123})\,,
\end{align}
where $\B{k}_{123} \equiv \B{k}_1+\B{k}_2+\B{k}_3$. We can trivially integrate over all amplitudes $A_{\B{k}}$
and by making use of the identity
\begin{align}
  \label{eq:PDF.int_amplitudes}
  {\cal I}(n) = \int_{0}^{\infty} A^n\,2A\,\text{e}^{-A^2}\,\D{A} = \Gamma\left(1+\frac{n}{2}\right)\,,
\end{align}
we see that both expressions above acquire a factor of ${\cal I}(1)^4 = (\sqrt{\pi}/2)^4$ (note that ${\cal
  I}(0) = 1$). Let us now consider the correlator
$\left<\epsilon_{\B{q}_1}\,\epsilon_{\B{q}_2}\,\epsilon_{\B{q}_3}\,\epsilon_{-\B{q}_{123}}\right>$ with
$\B{q}_1$, $\B{q}_2$, $\B{q}_3$ all in the upper half sphere, such that a contribution from
Eq.~(\ref{eq:PDF.4pQ1}) looks as follows:
\begin{align}
  \propto &\,\sum_{\substack{\B{k}_1,\,\B{k}_2,\,\B{k}_3 \\ \B{k}_i \neq \B{k}_j}}^{\text{uhs}}
  p^{(4)}(\B{k}_1,\,\B{k}_2,\,\B{k}_3,\,-\B{k}_{123}) \nonumber \\
  &\times\,\int \frac{\D{\theta_{\B{q}_1}}}{2\pi} \text{e}^{i\theta_{\B{q}_1}} \int
  \frac{\D{\theta_{\B{q}_2}}}{2\pi} \text{e}^{i\theta_{\B{q}_2}} \int \frac{\D{\theta_{\B{q}_3}}}{2\pi}
  \text{e}^{i\theta_{\B{q}_3}}
  \int \frac{\D{\theta_{\B{q}_{123}}}}{2\pi} \text{e}^{-i\theta_{\B{q}_{123}}} \nonumber \\
  &\times\,\prod_{\B{p} \neq \B{q}_i}^{\text{uhs}} \int \frac{\D{\theta_{\B{p}}}}{2\pi}
  \,\cos{\left(\theta_{\B{k}_1}+\theta_{\B{k}_2}+\theta_{\B{k}_3}-\theta_{\B{k}_{123}}\right)}\,.
\end{align}
As was already mentioned above, unless all phase factors are cancelled, the whole expression will evaluate to
zero. Non-zero contributions therefore stem from cases where each $\B{q}$-mode equals one of the $\B{k}$-modes,
giving in total $3! = 6$ different permutations. Eq.~(\ref{eq:PDF.4pQ2}) does not add to this exemplary
configuration and the full result is hence
\begin{align}
  \left<\epsilon_{\B{q}_1}\,\epsilon_{\B{q}_2}\,\epsilon_{\B{q}_3}\,\epsilon_{-\B{q}_{123}}\right> =
  \left(\frac{\sqrt{\pi}}{2}\right)^4 p^{(4)}(\B{q}_1,\,\B{q}_2,\,\B{q}_3,\,-\B{q}_{123})\,.
\end{align}
It can be checked that this holds true for all possible configurations of the $\B{q}$-modes.

\subsection{Six-point phase correlator}
\label{sec:six-point-phase}

The lowest order contributions to the connected six-point phase correlator come from terms with $n=4$ in the PDF
expansion. In this case we find three terms with six different phase factors,
\begin{align}
  \label{eq:PDF.p6Q1}
  {\cal Q}^{(4)}_1 =\,&\frac{1}{60} \sum_{\substack{\B{k}_1,\,\ldots,\,\B{k}_5 \\ \B{k}_i \neq \B{k}_j}}^{\text{uhs}}\hspace{-0.25em}
  A_{\B{k}_1}\,\cdots\,A_{\B{k}_5}\,A_{\B{k}_{12345}} \nonumber \\
  &\times\,\cos{\left(\theta_{\B{k}_1}+\theta_{\B{k}_2}+\theta_{\B{k}_3}+\theta_{\B{k}_4}+\theta_{\B{k}_5}-\theta_{\B{k}_{12345}}\right)} \nonumber \\
  &\times\,p^{(6)}(\B{k}_1,\,\B{k}_2,\,\B{k}_3,\,\B{k}_4,\,\B{k}_5,\,-\B{k}_{12345})\,,
\end{align}
\begin{align}
  \label{eq:PDF.p6Q2}
  {\cal Q}^{(4)}_2 =\,&\frac{1}{24} \sum_{\substack{\B{k}_1,\,\ldots,\,\B{k}_5 \\ \B{k}_i \neq \B{k}_j}}^{\text{uhs}}\hspace{-0.25em}
  A_{\B{k}_1}\,\cdots\,A_{\B{k}_5}\,A_{\B{k}_{12345}} \nonumber \\
  &\times\,\cos{\left(\theta_{\B{k}_1}+\theta_{\B{k}_2}+\theta_{\B{k}_3}+\theta_{\B{k}_4}-\theta_{\B{k}_5}-\theta_{\B{k}_{12345}}\right)} \nonumber \\
  &\times\,p^{(6)}(\B{k}_1,\,\B{k}_2,\,\B{k}_3,\,\B{k}_4,\,-\B{k}_5,\,-\B{k}_{12345})\,,
\end{align}
\begin{align}
  \label{eq:PDF.p6Q3}
  {\cal Q}^{(4)}_3 =\,&\frac{1}{36} \sum_{\substack{\B{k}_1,\,\ldots,\,\B{k}_5 \\ \B{k}_i \neq \B{k}_j}}^{\text{uhs}}\hspace{-0.25em}
  A_{\B{k}_1}\,\cdots\,A_{\B{k}_5}\,A_{\B{k}_{12345}} \nonumber \\
  &\times\,\cos{\left(\theta_{\B{k}_1}+\theta_{\B{k}_2}+\theta_{\B{k}_3}-\theta_{\B{k}_4}-\theta_{\B{k}_5}-\theta_{\B{k}_{12345}}\right)} \nonumber \\
  &\times\,p^{(6)}(\B{k}_1,\,\B{k}_2,\,\B{k}_3,\,-\B{k}_4,\,-\B{k}_5,\,-\B{k}_{12345})\,,
\end{align}
As for the the four-point phase correlator we first integrate out all amplitudes, which now gives rise to a
factor $(\sqrt{\pi}/2)^6$ each, where we have made use of Eq.~(\ref{eq:PDF.int_amplitudes}) again. We then
consider a six-point correlator of the form
$\left<\epsilon_{\B{q}_1}\,\epsilon_{\B{q}_2}\,\epsilon_{-\B{q}_{12}}\,\epsilon_{\B{q}'_1}\,\epsilon_{\B{q}'_2}\,\epsilon_{-\B{q}'_{12}}\right>$
and assume that $\B{q}_1$, $\B{q}_2$, $\B{q}'_1$, $\B{q}'_2 \in \text{uhs}$. In this case only
Eq.~(\ref{eq:PDF.p6Q2}) can give a non-vanishing contribution and for that $\B{q}_{12}$ and $\B{q}'_{12}$ must
equal either $\B{k}_5$ or $\B{k}_{12345}$ and all other $\B{q}$- and $\B{q}'$-modes must be identified with the
remaining $\B{k}$-modes. This gives $2 \times 4! = 48$ possibilities and thus we get
\begin{align}
  &\left<\epsilon_{\B{q}_1}\,\epsilon_{\B{q}_2}\,\epsilon_{-\B{q}_{12}}\,\epsilon_{\B{q}'_1}\,\epsilon_{\B{q}'_2}\,\epsilon_{-\B{q}'_{12}}\right>
  \nonumber \\ &= \left(\frac{\sqrt{\pi}}{2}\right)^6\,
  p^{(6)}(\B{q}_1,\,\B{q}_2,\,-\B{q}_{12},\,\B{q}'_1,\,\B{q}'_2,\,-\B{q}'_{12})\,,
\end{align}
which is also valid for any configuration of the six modes.

\subsection{Mixed five-point correlator}
\label{sec:mixed-five-point}

Finally, we have to consider the mixed five-point correlator between five phases and two amplitudes, whose
leading contributions are of the order $1/V^{3/2}$, i.e. $n = 3$. There are two terms in the Edgeworth
expansion, which have five different phase factors:
\begin{align}
  \label{eq:PDF.p5Q1}
  {\cal Q}^{(3)}_1 =\,&\frac{1}{12} \sum_{\substack{\B{k}_1,\,\ldots,\,\B{k}_4 \\ \B{k}_i \neq \B{k}_j}}^{\text{uhs}}\hspace{-0.25em}
  A_{\B{k}_1}\,\cdots\,A_{\B{k}_4}\,A_{\B{k}_{1234}} \nonumber \\
  &\times\,\cos{\left(\theta_{\B{k}_1}+\theta_{\B{k}_2}+\theta_{\B{k}_3}+\theta_{\B{k}_4}-\theta_{\B{k}_{1234}}\right)} \nonumber \\
  &\times\,p^{(5)}(\B{k}_1,\,\B{k}_2,\,\B{k}_3,\,\B{k}_4,\,-\B{k}_{1234})\,,  
\end{align}
\begin{align}
  \label{eq:PDF.p5Q2}
  {\cal Q}^{(3)}_2 =\,&\frac{1}{6} \sum_{\substack{\B{k}_1,\,\ldots,\,\B{k}_4 \\ \B{k}_i \neq \B{k}_j}}^{\text{uhs}}\hspace{-0.25em}
  A_{\B{k}_1}\,\cdots\,A_{\B{k}_4}\,A_{\B{k}_{1234}} \nonumber \\
  &\times\,\cos{\left(\theta_{\B{k}_1}+\theta_{\B{k}_2}+\theta_{\B{k}_3}-\theta_{\B{k}_4}-\theta_{\B{k}_{1234}}\right)} \nonumber \\
  &\times\,p^{(5)}(\B{k}_1,\,\B{k}_2,\,\B{k}_3,\,-\B{k}_4,\,-\B{k}_{1234})\,,  
\end{align}
Let us consider the correlator
$\left<\delta_{\B{k}_1}\,\delta_{\B{k}_2}\,\epsilon_{\B{q}_1}\,\epsilon_{\B{q}_2}\,\epsilon_{\B{q}_3}\right>$,
where we assume that $\B{k}_1,\,\B{k}_2,\,\B{q}_1,\,\B{q}_2 \in \text{uhs}$ and $\B{q}_3 \in \text{lhs}$. All
terms involving ${\cal Q}_2^{(3)}$ will evaluate to zero and from the amplitude integrals we obtain a factor
${\cal I}(1)^3\times{\cal I}(2)^2 = (\sqrt{\pi}/2)^3$. We are thus left with the following phase integrals:
\begin{align}
  \left<\delta_{\B{k}_1}\,\delta_{\B{k}_2}\,\epsilon_{\B{q}_1}\,\epsilon_{\B{q}_2}\,\epsilon_{\B{q}_3}\right>
  =\,& \sqrt{P(k_1)\,P(k_2)}\,\left(\frac{\sqrt{\pi}}{2}\right)^3 \frac{V}{12}\,\int\hspace{-0.5em}\prod_{\B{p}
    \in \text{uhs}}\frac{\D{\theta_{\B{p}}}}{2\pi} \nonumber \\
  &\hspace{-4.25em}\times \hspace{-0.5em} \sum_{\substack{\B{u}_1,\,\ldots,\,\B{u}_4 \in \text{uhs} \\ \B{u}_i \neq \B{u}_j}}
  \hspace{-0.5em} \cos{\left(\theta_{\B{u}_1}+ \cdots + \theta_{\B{u}_4}-\theta_{\B{u}_{1234}}\right)} \nonumber
  \\
  &\hspace{-4.25em}\times\,p^{(5)}(\B{u}_1,\ldots,-\B{u}_{1234})\, 
  \text{e}^{i\left(\theta_{\B{k}_1}+\B{k}_2+\B{q}_2+\B{q}_2-\theta_{-\B{q}_3}\right)}\,,
\end{align}
which only give a non-vanishing result if we impose the condition that $\B{q}_3 = -\B{k}_1 - \B{k}_2 - \B{q}_1
- \B{q}_2$. We then have $4! = 24$ possibilities of matching the various $\B{k}$- and $\B{q}$-modes with the
$\B{u}$-vectors and after taking the continuum limit we finally get:
\begin{align}
  \left<\delta_{\B{k}_1}\,\delta_{\B{k}_2}\,\epsilon_{\B{q}_1}\,\epsilon_{\B{q}_2}\,\epsilon_{\B{q}_3}\right> =
  &\,(2\pi)^3\,\left(\frac{\sqrt{\pi}}{2}\right)^3\,\sqrt{P(k_1)\,P(k_2)} \nonumber \\
  &\times\,p^{(5)}(\B{k}_1,\,\B{k}_2,\,\B{q}_1,\,\B{q}_2,\,\B{q}_3) \nonumber \\
  &\times\,\delta_D\left(\B{k}_1+\B{k}_2+\B{q}_1+\B{q}_2+\B{q}_3\right)\,.
\end{align}
As before this result is not restricted to the particular configuration of wavevectors we have chosen above.

\section{Galaxy power spectrum at one-loop order}
\label{sec:galaxy-powerspectrum}

At linear order the power spectrum is only dependent on the single bias parameter $b_1$, while at the one-loop
level non-linear and non-local bias introduce some additional terms that we need to account for. The full galaxy
power spectrum at one-loop order is therefore given by \citep{McDonald2009}:
\begin{align}
  \label{eq:galaxy-powerspectrum.Pg}
  P_g(k) =\,& b_1^2\,P(k) + 2b_2b_1\,P_{b2}(k) + 2b_{s^2}b_1\,P_{bs2}(k) \nonumber \\ 
  &+ b_2^2\,P_{b22}(k) + 2b_2b_{s^2}\,P_{b2,bs2}(k) + b_{s^2}^2\,P_{bs22}(k) \nonumber \\ &+ 2b_1b_{3\text{nl}}\,\sigma_3^2(k)\,P_L(k)\,,
\end{align}
where $P(k)$ and $P_L(k)$ denote the one-loop and linear dark matter power spectra, respectively. The power
spectra that appear in combination with the bias parameters $b_2$ and $b_{s^2}$ are given by the following
integrals,
\begin{align}
  P_{b2}(k) = \int \frac{\D{^3q}}{(2\pi)^3} P_L(q)\,P_L(|\B{k}-\B{q}|)\,F_2(\B{q},\,\B{k}-\B{q})\,,
\end{align}
\begin{align}
  P_{bs2}(k) =\,&\int \frac{\D{^3q}}{(2\pi)^3}P_L(q)\,P_L(|\B{k}-\B{q}|)\,F_2(\B{q},\,\B{k}-\B{q}) \nonumber \\ 
  &\times\,S_2(\B{q},\,\B{k}-\B{q})\,,
\end{align}
\begin{align}
  P_{b2,bs2}(k) =\,&-\frac{1}{2}\int \frac{\D{^3q}}{(2\pi)^3}
  P_L(q)\,\Bigg[\frac{2}{3}P_L(q)\Bigg. \nonumber \\ &\Bigg.-P_L(|\B{k}-\B{q}|)\,S_2(\B{q},\,\B{k}-\B{q})\Bigg]\,,
\end{align}
\begin{align}
  P_{bs22}(k) =\,&-\frac{1}{2}\int \frac{\D{^3q}}{(2\pi)^3}
  P_L(q)\,\Bigg[\frac{4}{9}P_L(q)\Bigg. \nonumber \\ &\Bigg.-P_L(|\B{k}-\B{q}|)\,S_2(\B{q},\,\B{k}-\B{q})^2\Bigg]\,,
\end{align}
\begin{align}
  P_{b22}(k) =\,&-\frac{1}{2}\int \frac{\D{^3q}}{(2\pi)^3}P_L(q)\,\left[P_L(q)-P_L(|\B{k}-\B{q}|)\right]\,,
\end{align}
\begin{align}
  \sigma_3^2(k) =\,&-\frac{1}{2}\int \frac{\D{^3q}}{(2\pi)^3}
  P_L(q)\,\Bigg[\frac{5}{6} + \frac{15}{8}S_2(\B{q},\,\B{k}-\B{q})\,S_2(-\B{q},\,\B{k})\Bigg. \nonumber \\ 
  &\Bigg.-\frac{5}{4}S_2(\B{q},\,\B{k}-\B{q})\Bigg]\,,
\end{align}
where the kernel functions $F_2$ and $S_2$ are defined in Eqs.~(\ref{eq:predictions.F2}) and
(\ref{eq:predictions.S2}). Assuming that galaxy bias is local in Lagrangian space, the non-local terms can be
related at first order to the linear bias term as follows \citep{Baldauf2012, Chan2012, Saito2014}:
\begin{align}
  &b_{s^2} = -\frac{4}{7}\,(b_1-1)\,, \\
  &b_{3\text{nl}} = \frac{32}{315}\,(b_1-1)\,.
\end{align}


\bsp	
\label{lastpage}
\end{document}